\documentclass[pdftex,reprint,nofootinbib,showpacs,showkeys,prd,floatfix]{revtex4-2}

\usepackage[american]{babel}

\newcommand{\phys}{{\textnormal{phys}}}  
\newcommand{\nul}{{\scriptscriptstyle(0)}}
\newcommand{\een}{{\scriptscriptstyle(1)}}
\renewcommand{\S}{\textsection}
\newcommand{\dif}{\mathrm{d}}
\newcommand{\apjl}{Astrophys. J. Lett.}
\newcommand{\physrep}{Phys. Rep.}

\usepackage{amsmath}
\usepackage{amssymb}
\usepackage{graphicx}
\usepackage{bm}

\usepackage[pdftitle={Earliest Structures in the Universe can be
  explained by a Relativistic Cosmological Perturbation
  Theory}, pdfauthor={Pieter G. Miedema}, pdfsubject={Cosmology},
pdfstartview={FitH}, pdfkeywords={Cosmology, Early Universe, Structure
  Formation, Perturbation Theory, Gauge Problem}, bookmarks,
bookmarksnumbered, colorlinks, linkcolor={blue},
citecolor={blue}]{hyperref}

\begin{document}


\title{Earliest Structures in the Universe can be explained \protect\\
  by a Relativistic Cosmological Perturbation Theory}



\author{Pieter G.\ Miedema}
\email[]{pg.miedema@protonmail.com}
\affiliation{Independent Researcher\\ Breda, The Netherlands}



\date{April 1, 2026}

\begin{abstract}
  A relativistic cosmological perturbation theory for the
  Friedmann-Lemaître-Robertson-Walker universe is presented that
  explains the masses and formation times of the first structures in
  our universe.  First, it is shown that, without a coordinate system
  being used, quantities intended to represent energy density and
  particle number density perturbations can be defined in only one
  way.  The Newtonian limit, where the pressure becomes zero, proves
  that these quantities are indeed the perturbations of the energy
  density and the particle number density.  Then, after selecting a
  reference frame, a perturbation theory will be formulated based on
  these quantities.  This formulation considers the local perturbation
  to the spatial curvature resulting from a density perturbation, the
  local fluid velocity due to pressure gradients caused by the
  self-gravity of the density perturbation, and entropy perturbations,
  all of which are necessary for structure formation.  Pressure
  perturbations consist of two components: an adiabatic component
  caused by the density perturbation itself, and a random,
  nonadiabatic component resulting from the rapid, chaotic transition
  to the era when matter and radiation were decoupled. Immediately
  after decoupling, negative nonadiabatic pressure perturbations in
  various density perturbations enabled their rapid growth over a
  short period of time.  This brief period ended when the total
  pressure perturbation became positive. Subsequently, the density
  perturbations gradually grew toward their nonlinear phase, which was
  reached early~on.
\end{abstract}


\maketitle



\section{Introduction}
\label{sec:introduction}

Structures in our universe, such as stars, galaxies, and black holes,
form through the gravitational growth of small density perturbations.
Therefore, it is reasonable to expect that a cosmological perturbation
theory based on general relativity --- the most accurate theory of
gravity to date --- would explain the masses and formation times of
the first structures.  There are two main reasons why such a theory
does not yet exist. First, the \emph{gauge problem of cosmology} has
not yet been solved.  Second, particle number density has not been
considered, so the pressure gradients necessary for fluid flow ---
which is required for structure formation after matter and radiation
decoupled --- cannot be taken into account.  This article resolves
both issues.

The study of the evolution of density perturbations by way of
linearization of the Einstein equations and conservation laws with
respect to a flat Friedmann-Lemaître-Robertson-Walker (\textsc{flrw})
universe was pioneered by Lifshitz~\cite{lifshitz1946} and Lifshitz \&
Khalatnikov~\cite{c15}.  They observed that solutions to the
linearized Einstein equations and conservation laws are dependent on
the choice of a coordinate system, i.e., the solutions are
\emph{gauge-dependent}.  Consequently, these solutions can be changed
by linear coordinate transformations, also known as \emph{gauge
  transformations}. This means that these solutions have no physical
meaning, which is a problem known as the \emph{gauge problem of
  cosmology}.  Solving this long-standing problem is the only way to
obtain physically significant solutions from the linearized Einstein
equations and conservation laws.  Doing so is paramount to explaining
how the earliest
structures~\cite{2025ApJ...994L..11N,2023ARA&A..61...65K} in the
universe formed from small density perturbations.

Physical phenomena such as density, temperature, pressure, and entropy
perturbations do not depend on the choice of a coordinate system.
Accordingly, the corresponding mathematical quantities describing
these phenomena must be defined without using a reference frame.  In
other words, these quantities must be \emph{gauge-invariant}.

Two different pioneering methods to address the gauge problem of
cosmology have been proposed in the literature.  In his seminal
article, Bardeen\,\cite{c13} introduced gauge-invariant quantities
intended to represent energy density perturbations as linear
combinations of gauge-dependent quantities.  Building on Bardeen's
work, Mukhanov\,\textit{et~al.}\,\cite{mfb1992,Mukhanov-2005} made
significant contributions to the subject.
Ellis\,\textit{et~al.}\,\cite{Ellis1,Ellis2,Bruni:1992dg} present a
second method that takes advantage of the fact that the gradient of a
gauge-dependent quantity is gauge-invariant.  For a comprehensive
overview of this approach and its further extension, see
Tsagas\,\textit{et al.}\,\cite{2008PhR...465...61T}.

Unfortunately, these approaches have some disadvantages.
Firstly, they allow for different definitions of gauge-invariant
quantities.  Since these approaches lack a Newtonian limit, it is
impossible to determine which gauge-invariant quantity corresponds to
an energy density perturbation.  Secondly, these approaches do not
account for the particle number density.  Thirdly, metric
perturbations are explicitly present in the evolution equations for
density perturbations.  Finally, these approaches have not been
extended to open and closed \textsc{flrw} universes.

Many new ideas have been added to this field.  See Ellis's $2017$
review article~\cite{Ellis2017} and other publications, e.g.,
Refs.\,\cite{kodama1984,1997astro.ph..1137H,
  ellis-1998,2000PhRvD..62d3527W,2001PhLA..292..173K,2005CQGra..22.3181N,
  2005AcPPB..36.2133G,
  Malik:2008im,Uggla_2011,Ellis-Maartens-MacCallum-2012,
  2012arXiv1208.5931K, 2013GReGr..45.1989M,
  2014EPJC...74.2917A,2018CQGra..35o5012G,Giesel_2018, Di_Gioia_2019,
  2020PhRvD.102b3524G, 2023CQGra..40u5006F,2023PhRvD.107j3510J}.
These researchers proposed alternative perturbation theories using
different gauge-invariant quantities that differ from those used by
Bardeen and Mukhanov\,\textit{et~al\@.} However, the most fundamental
question regarding which gauge-invariant quantity accurately and
unambiguously represents a density perturbation remains unanswered.

This article presents a relativistic cosmological perturbation theory
that does not have the aforementioned disadvantages.  Firstly, it
demonstrates that a gauge-invariant quantity describing a density
perturbation can be defined in precisely one way.  Secondly, in
addition to energy density, particle number
density~\cite{2001CQGra..18.3917M} is considered.  Both are crucial
for structure formation.  Finally, it is shown that perturbations in
the spatial metric and its derivatives are included in the
perturbations of the spatial Ricci scalar and expansion scalar.  The
use of these quantities significantly simplifies the system of
linearized Einstein equations and conservation laws, leading not only
to a perturbation theory that encompasses closed, flat, and open
\textsc{flrw} universes but also allowing the Newtonian limit to be
derived.

Unlike the approaches in
Refs.\,\cite{Liu_2022,sharma2024originsupermassiveblackholes,
  Zhoolideh_Haghighi_2024,2024ApJ...976...13M,
  Bird_2024,dolgov2025tensionhstjwstlambdacdmcosmology}, this article
does not introduce any ad hoc modifications to the theory of
relativity, free parameters, Cold Dark Matter (\textsc{cdm}), or the
peculiar
velocities~\cite{tsagas2025largescalepeculiarvelocitiesuniverse,
  tsagas2025generalrelativityearlygalaxy,2026arXiv260121741P} of
matter.  However, the linearized Einstein equations and conservation
laws naturally encompass the velocity of matter caused by pressure
gradients, as well as the perturbed Ricci scalar caused by a density
perturbation. These must be taken into account.

\section{Development and Overview of the Perturbation Theory}

Sec.\,\ref{sec:cosmic-fluid} explains why, in addition to energy
density, particle number density must also be considered.

Sec.\,\ref{sec:overview-gauge-problem} explains the gauge problem in
cosmology and provides a solution by demonstrating that a
gauge-invariant quantity that describes energy density perturbations
can be defined in only one way. The same applies to the
gauge-invariant perturbation of the particle number density.  These
definitions imply that the gauge-invariant perturbation to the
expansion of the universe is equal to zero.  In
Sec.\,\ref{sec:newt-limit}, the Newtonian limit will be used to
demonstrate that these quantities represent density perturbations.

To formulate a perturbation theory based on these quantities, a
coordinate system is necessary.  The reasons why a synchronous
reference frame is the best choice for this task are explained in
Sec.\,\ref{subsec:Sys-of-Ref}\@.

In Sec.\,\ref{sec:basic-equations}, the Einstein equations and
conservation laws for \textsc{flrw} universes, as well as their
linearized counterparts, are presented with respect to synchronous
coordinates.  Given that the evolution of density perturbations is
considerably influenced by pressure perturbations, the covariant
divergence of the fluid velocity, and the perturbation of the Ricci
scalar, expressions are derived for these quantities.

Since we are working in synchronous coordinates, we can apply the
decomposition theorem for symmetric, second-rank spatial tensors, such
as the perturbed metric tensor and the perturbed spatial Ricci tensor,
as well as the Helmholtz decomposition for the fluid velocity vector.
These theorems are discussed in Sec.\,\ref{sec:decomp-h-u}.  According
to these theorems, the linearized Einstein equations and conservation
laws decompose into three independent systems of equations: one for
tensor perturbations, one for vector perturbations, and one for scalar
perturbations.  In the literature, the decomposition theorem is only
applied to the spatial metric tensor perturbation. However, this is
insufficient.  Since the perturbed Ricci scalar plays an important
role in the evolution of density perturbations and in the Newtonian
limit, the decomposition theorem must also be applied to the perturbed
Ricci tensor.

Sec.\,\ref{subsec:decomp-lin-eqs} uses these theorems to show that
only scalar perturbations are associated with density perturbations.
In this case, the perturbed Ricci scalar contains the spatial
derivatives of two independent potentials, while the perturbed
expansion scalar contains their time derivatives. In the Newtonian
limit, one of the potentials becomes the Newtonian potential while the
other becomes irrelevant.

Since we are interested in density perturbations, we derive in
Sec.\,\ref{sec:scalar-pert} a system of equations describing the
evolution of scalar perturbations.  This system contains three
conservation laws: one for the energy density perturbation, one for
the particle number density perturbation, and one for the momentum.
It also contains two constraint equations. The algebraic energy
density constraint equation contains the perturbed Ricci scalar, and
the momentum constraint equation describes how the perturbed Ricci
scalar evolves.

The Newtonian limit is reached when the velocities of the particles
become negligible compared to the speed of light --- that is, when the
pressure approaches zero.  Using this limit it is shown in
Sec.\,\ref{sec:newt-limit} that the newly defined gauge-invariant
quantities in Sec.\,\ref{sec:overview-gauge-problem} represent
perturbations of the energy and particle number densities.  In other
words, in the Newtonian limit, the momentum constraint equation
ensures that the Newtonian potential and the energy density
perturbation are independent of time. It also makes sure that the
energy density constraint equation is identical to the Poisson
equation of Newtonian gravity.  Additionally, the well-known special
relativistic equation relating rest mass density to energy density is
obtained.

The Newtonian limit can only be obtained using the reformulated
evolution equations for scalar perturbations given in
Sec.\,\ref{sec:scalar-pert}\@.  Since these equations are not
presented in this form in the literature, previous cosmological
perturbation theories have no Newtonian limit.  Without a clearly
defined Newtonian limit, it is impossible to demonstrate which
gauge-invariant quantity represents a density perturbation. This
article resolves this issue.

The Newtonian limit demonstrates furthermore that relativistic gauge
transformations become Newtonian gauge transformations.  Consequently,
the conventional evolution equation for relative perturbations in a
flat \textsc{flrw} universe filled with a nonrelativistic fluid ---
whether derived from general relativity or Newtonian gravity adapted
to the expansion of the universe --- also has gauge-dependent
solutions devoid of physical meaning.  Consequently, Newtonian gravity
is not suitable for studying density perturbations in the universe,
regardless of their scale.  This means that no distinction can be made
between density perturbations inside and outside the horizon.

The perturbation theory is presented in Sec.\,\ref{sec:flat-pert}\@.
It describes the evolution of density perturbations in closed, flat,
and open \textsc{flrw} universes. The theory incorporates the
perturbed Ricci scalar, the pressure, the entropy, and the covariant
divergence of the fluid velocity.  This is demonstrated by the
algorithm in Appendix\;\ref{app:computer-algebra}.

In Sec.\,\ref{sec:thermodyn}, expressions are derived for the
perturbations of entropy, temperature, and pressure to connect the
perturbation theory to an equation of state for a cosmological fluid.
It will be shown that pressure perturbations have two components: an
adiabatic component caused by the density perturbation itself, and a
nonadiabatic component.

In Sec.\,\ref{sec:flat-flrw-universe}, we apply the perturbation
theory to a flat \textsc{flrw} universe in its radiation-dominated
phase and after the decoupling of matter and radiation.

Finally, we summarize the results and draw conclusions in
Sec.\,\ref{sec:conclusion}.

The article is accompanied by three appendices.
Appendix\;\ref{app:computer-algebra} presents an algorithm that, when
executed with the assistance of a computer algebra program, derives
the perturbation theory from the background Einstein equations and the
system of equations for scalar perturbations.  Since the latter system
is a new result, its correctness is demonstrated in
Appendices\;\ref{app:consistency-check}
and\;\ref{sec:stand-theory}. Finally, Appendix\;\ref{sec:stand-theory}
compares the conventional perturbation equations with the perturbation
theory developed in this article.

\section{Equations of State}
\label{sec:cosmic-fluid}

We distinguish between two main eras of the universe: the period
before the decoupling of matter and radiation and the post-decoupling
era.

Before decoupling, the cosmological fluid can be treated as a
relativistic fluid consisting of radiation, ordinary matter, and
\textsc{cdm}, where the contributions of matter and \textsc{cdm} to
the pressure are negligible and the radiation pressure is of the same
order of magnitude as the energy density.  See, e.g.,
Ref.\,\cite{mfb1992}, Eq.\,(5.49), Ref.\,\cite{Mukhanov-2005},
Eq.\,(1.78), and Ref.\,\cite{kodama1984}, {\S\,V-1}.

The conditions in the universe became significantly different after
decoupling of matter and radiation.  This is because the pressure in
the cosmological fluid has become negligible compared to its energy
density.  Therefore, pressure can be neglected in the Einstein
equations and conservation laws describing the evolution of a
\textsc{flrw} universe.  However, pressure \emph{perturbations} cannot
be ignored when studying structure formation.  The self-gravity of a
local density perturbation creates pressure gradients.  These
gradients then cause particle flows, thereby increasing a local
density perturbation.  The linearized momentum conservation law shows
that the absence of pressure perturbations implies an absence of fluid
flows.  This results in \emph{static} density perturbations.
Therefore, pressure perturbations are crucial to understanding how
structures in the universe formed.  After decoupling, the cosmological
fluid can be treated as a perfect nonrelativistic gas.  In this case,
pressure depends on the average kinetic energy density of the
particles.  Therefore, it is essential to include, next to the rest
energy density, also the kinetic energy density in the analysis.  Both
pressure and kinetic energy density depend on the particle number
density and temperature.  Because of this, particle number density and
temperature must be included in the equations of state.  See
Ref.\,\cite{c8}, Sec.\,2.10 on relativistic hydrodynamics, and
Refs.\,\cite{2001CQGra..18.3917M,mg16-2021}. This results in random,
nonadiabatic density perturbations, which are crucial for the
formation of structures in the universe.

The aforementioned considerations are consistent with thermodynamic
principles, namely that energy density~$\varepsilon$ and pressure~$p$
are functions of particle number density~$n$ and temperature~$T$,
i.e.,
\begin{equation}
  \label{eq:e-p-dependend-n-T}
      \varepsilon=\varepsilon(n,T), \quad p=p(n,T).
\end{equation}
In order to facilitate the calculations, it is advantageous to
eliminate the temperature from the given equations of state.  This
yields the equation of state for the pressure:
\begin{equation}
  \label{eq:equation-of-state-pressure}
  p=p(n,\varepsilon),
\end{equation}
which will be used in the subsequent analysis.

\section{Gauge Problem and its Solution}
\label{sec:overview-gauge-problem}

Einstein's equations and conservation laws are invariant under
coordinate transformations $x^\mu\rightarrow
x^{\prime\mu}(x^\nu)$. This signifies that there are no preferred
coordinate systems. Consequently, the linearized Einstein equations
and conservation laws are \emph{gauge-invariant}, meaning they are
invariant under linear, infinitesimal space-time transformations, also
known as \emph{gauge transformations}, which are given by
\begin{equation}
     x^\mu \rightarrow x^{\prime\mu}=x^\mu - \xi^\mu(t,\bm{x}), \label{func}
\end{equation}
where the gauge functions $\xi^\mu(t,\bm{x})$ are four arbitrary
infinitesimal functions of time, $x^0=ct$ with $c$ the speed of light,
and space, $\bm{x}=(x^1,x^2,x^3)$, coordinates. The gauge problem of
cosmology stems from the linear transformation~(\ref{func}) and the
linearity of the linearized Einstein equations and conservation laws.
This problem will now be examined.

Consider a closed, flat, or open \textsc{flrw} universe.  The
evolution equations for these universes contain three scalars: the
energy density $\varepsilon$, the particle number density $n$, and the
expansion~$\theta$. These scalars are defined as follows
\begin{equation}
  \label{eq:scalars-flrw}
    \varepsilon := T^{\mu\nu} u_\mu u_\nu, \quad
    n := N^\mu u_\mu, \quad
    \theta := u^\mu{}_{;\mu},
\end{equation}
where $u^\mu$ is the fluid four-velocity normalized to unity,
${u^\mu u_\mu=1}$, and $N^\mu:= nu^\mu$ represents the cosmological
particle current four-vector, which satisfies the particle number
conservation law ${N^\mu{}_{;\mu}=0}$.  A semicolon denotes covariant
differentiation.  Let $S_\nul(t)$ denote the quantities
$\varepsilon_\nul$, $n_\nul$, and $\theta_\nul$. These quantities
satisfy the background equations, that is to say, the unperturbed
Einstein equations and conservation laws.  Let $S_\een(t,\bm{x})$ be
their perturbed counterparts $\varepsilon_\een$, $n_\een$,
and~$\theta_\een$, which satisfy the linearized equations. As a
consequence of the linearity of these equations, new solutions that
have precisely the same physical content can be generated. These new
solutions are given by (see Weinberg~\cite{c8}, Sec.\,10.9 for a
detailed explanation)
\begin{equation}
\label{eq:gauge-problem}
S^\prime_\een=S_\een+\mathcal{L}_\xi S_\nul = S_\een+\xi^0\dot{S}_\nul,
\quad S=\varepsilon, n, \theta.
\end{equation}
In this expression, the operator $\mathcal{L}_\xi$ is the Lie
derivative with respect to the infinitesimal four-vector~$\xi^{\mu}$.
An overdot denotes differentiation with respect to $x^{0}=ct$.  The
solutions $S_\een$ and $S^\prime_\een$ contain the so-called
\emph{gauge modes} which are given by
$\hat{S}_\een=\xi^0\dot{S}_\nul$.  This indicates that the solutions
$\varepsilon_\een$, $n_\een$, and $\theta_\een$ to the linearized
equations are \emph{gauge-dependent}. Since these solutions depend on
the choice of a reference frame, they are devoid of physical
significance. Therefore, $\varepsilon_\een$, $n_\een$,
and~$\theta_\een$ do not represent physical phenomena.  This is the
\emph{gauge problem of cosmology}, whereby the coordinate artifacts
and the underlying physics are inextricably linked in the solution to
the linearized equations.  Now that the gauge problem of cosmology has
been understood, a solution can be found.

The physics of the perturbations is hidden in the solutions to the
linearized Einstein equations and conservation laws. Consequently,
quantities that describe physical phenomena must be linear
combinations of gauge-dependent solutions to these equations. Since
the transformation of the quantities $\varepsilon_\een$, $n_\een$,
and~$\theta_\een$ under the infinitesimal transformation~(\ref{func})
is identical to that described by~(\ref{eq:gauge-problem}), we can
construct gauge-invariant quantities by combining these three
gauge-dependent quantities in a way that eliminates the gauge
modes~$\xi^0\dot{S}_\nul$.  This results in three distinct sets of
linear combinations. In each of these sets, precisely one
gauge-invariant quantity is identically zero. Since we are only
interested in energy density and particle number density
perturbations, the only possible set is
\begin{subequations}
  \label{subeq:gi-quant}
\begin{align}
   & \varepsilon_\een^\phys
        :=\varepsilon_\een-\dfrac{\dot{\varepsilon}_\nul}
        {\dot{\theta}_\nul}\theta_\een, \quad
   n_\een^\phys:= n_\een-\dfrac{\dot{n}_\nul}
        {\dot{\theta}_\nul}\theta_\een,   \label{e-n-gi} \\
      &   \theta_\een^\phys :=\theta_\een-\dfrac{\dot{\theta}_\nul}
        {\dot{\theta}_\nul}\theta_\een=0. \label{theta-gi}
\end{align}
\end{subequations}
From these definitions, it can be inferred that the general form of a
gauge-invariant quantity describing a physical phenomenon is given~by
\begin{equation}
  \label{eq:alg-def-gi}
     q^{\phys}_{\een}:=q_{\een}-
         \dfrac{\dot{q}_{\nul}}{\dot{\theta}_{\nul}}\theta_{\een},
\end{equation}
where $q$ stands for energy density~$\varepsilon$, particle number
density~$n$, temperature~$T$, pressure~$p$, or the entropy per
particle~$s$. 

The quantities~(\ref{e-n-gi}) represent density perturbations
and~(\ref{theta-gi}) represents the perturbation to the expansion of
the universe. This will be demonstrated in Sec.\,\ref{sec:newt-limit}
on the Newtonian limit.

\section{Selection of a Reference Frame}
\label{subsec:Sys-of-Ref}

The quantities $\varepsilon^\phys_\een$ and $n^\phys_\een$ are defined
without the use of a coordinate system.  However, to derive evolution
equations for these quantities, it is first necessary to choose a
suitable reference frame.  Since general relativity is covariant, and
the quantities in question are gauge-invariant, it follows that the
evolution equations for these quantities can be derived in any chosen
coordinate system.  The selection of an appropriate coordinate system
for this problem is dependent on two criteria, which are outlined
below.

Firstly, to show that~$\varepsilon^\phys_\een$ and~$n^\phys_\een$
represent density perturbations requires taking the Newtonian limit.
In the context of general relativity, coordinate transformations are
in general space-time transformations, specifically
$x^{\mu}\rightarrow x^{\prime\mu}(x^{\nu})$.  In the linear case,
space-time transformations are given by~(\ref{func}).  In Newtonian
gravity, space and time are different and independent
entities. Consequently, in the Newtonian limit, the relativistic
space-time transformations~(\ref{func}) must be automatically
separated into independent space and time transformations.  Secondly,
it would be preferable to have a coordinate system that would
facilitate the derivation of the perturbation theory.  It will now be
demonstrated that both of these requirements can be met by employing
the same coordinate system.

Firstly, in Newtonian gravity, where space and time are treated as
distinct and independent entities, all coordinate systems are
inherently synchronous.  In view of the Newtonian limit, it can be
seen that a synchronous reference frame represents the optimal choice.
In these coordinates, the metric tensor $g_{\mu\nu}(t,\bm{x})$ of the
\textsc{flrw} universes is given by
\begin{equation}
  \label{eq:metric-flrw}
 g_{00}=1, \quad g_{0i}=0, \quad g_{ij}=   -a^{2}(t)\tilde{g}_{ij}(\bm{x}).
\end{equation}
The scale factor of the universe is designated as $a(t)$.  Given that
$g_{00}=1$, coordinate time is equal to proper time.  The
synchronicity condition is $g_{0i}=0$, as explained in Ref.\,\cite{c8},
Sec.\,11.8 and Ref.\,\cite{I.12}, \S\,84. The metric tensor of the
three-dimensional maximally symmetric subspaces of constant time is
denoted by~$\tilde{g}_{ij}$.  The Killing equations
$\xi_{\mu;\nu}+\xi_{\nu;\mu}=0$ and~(\ref{eq:metric-flrw}) demonstrate
that the gauge functions $\xi^{\mu}(t,\bm{x})$ in the
transformation~(\ref{func}) become
\begin{equation}
  \label{eq:synchronous}
\xi^0=\psi(\bm{x}), \quad
\xi^i=\tilde{g}^{ik}(\bm{x})\dfrac{\partial\psi(\bm{x})}{\partial x^k}
       \int\!\! \frac{{\dif}t}{a^2(t)} + \chi^i(\bm{x}),
\end{equation}
if only transformations between synchronous coordinates are allowed
(see Ref.\,\cite{c8}, Sec.\,10.9).  In
expressions~(\ref{eq:synchronous}), $\psi(\bm{x})$ and
$\chi^i(\bm{x})$ are four arbitrary infinitesimal functions of the
spatial coordinates.

Secondly, in synchronous coordinates the space-space components
$R_{ij}$ of the four-dimensional Ricci tensor $R_{\mu\nu}$ are
partitioned into two distinct parts.  One part exclusively comprises
the time derivatives of the spatial metric tensor $g_{ij}$ while the
second part is the Ricci tensor~${^3\!}R_{ij}$ of the
three-dimensional subspaces.  This is demonstrated in
Ref.\,\cite{I.12}, \S\,97.  This implies that the spatial metric and
Ricci tensor perturbations are both tensors.  Furthermore, the spatial
part of the fluid four-vector is a vector in synchronous coordinates.
This allows us to apply the decomposition theorem for second-rank,
symmetric three-tensors and the Helmholtz theorem to decompose the
fluid vector.  Using these theorems, it is shown that the linearized
Einstein equations and conservation laws can be decomposed into three
mutually exclusive systems: one for tensor perturbations, one for
vector perturbations, and one for scalar perturbations.  Next, it will
be demonstrated that only scalar perturbations are associated with
density perturbations. Therefore, when studying density perturbations,
it is sufficient to consider the evolution equations for scalar
perturbations.  This system is easier to use than the original set of
equations, thus facilitating the derivation of a cosmological
perturbation theory. Because synchronous coordinates are also
compatible with Newtonian gravity, they will be used from this point
onward.

\section{Einstein Equations and Conservation Laws}
\label{sec:basic-equations}

As was concluded in the previous section, synchronous coordinates are
the optimal choice for deriving a perturbation theory.  This section
presents the background Einstein equations and conservation laws, as
well as their linearized counterparts, in this coordinate system.

The background energy density constraint equation and its linearized
counterpart are expressed in a manner consistent with the contracted
Bianchi identities. These identities state that the Riemann tensor
satisfies the equation $R^{\mu}{}_{\nu;\mu}=\tfrac{1}{2}R_{,\nu}$, where a comma denotes ordinary differentiation.
This can be expressed in an equivalent manner as the four-divergence
of the Einstein tensor, i.e., ${G^{\mu\nu}{}_{;\nu}=0}$.  This result
demonstrates that the constraint equations contain at most first-order
time derivatives of the metric, which is precisely what is required
for the development of a cosmological perturbation theory.  For
further details see Ref.\,\cite{c8}, Secs.\,6.8 and~7.5 and
Ref.\,\cite{I.12}, \S\,92 and~\S\,95.

\subsection{Background Equations}

\label{sec:background-einstein-equations}

The system of background Einstein equations and conservation laws for
closed, flat, and open \textsc{flrw} universes filled with a perfect
fluid with an energy-momentum tensor as defined by
\begin{equation}
  \label{eq:en-mom-tensor}
  T^{\mu\nu}=(\varepsilon+p)u^\mu u^\nu - p g^{\mu\nu}, \quad 
p=p(n,\varepsilon),
\end{equation}
is represented by the following equations:
\begin{subequations}
\label{subeq:einstein-flrw}
\begin{alignat}{3}
  3H^2 & =\tfrac{1}{2}\,{^3\!}R_\nul+\kappa\varepsilon_\nul+\Lambda,
         \quad & \kappa& = 8\pi G_{\text{N}}/c^4  & \label{FRW3}\\
  {^3\!}\dot{R}_\nul & =-2H\,{^3\!}R_\nul, & & \label{FRW3a}\\
  \dot{\varepsilon}_\nul & = -3H\varepsilon_\nul(1+w), & 
           w & := p_\nul/\varepsilon_\nul, & \label{FRW2} \\
  \dot{n}_\nul & = -3Hn_\nul. & &  \label{FRW2a}
\end{alignat}
\end{subequations}
The $G_{0i}$ constraint equations and the $G_{ij}$ dynamical equations
with $i\neq j$ are satisfied identically. The $G_{ii}$ dynamical
equations are identical to the time derivative of the $G_{00}$
constraint equation (Friedmann equation)~(\ref{FRW3}). Therefore, the
$G_{0i}$ and $G_{ij}$ equations can be omitted. The cosmological
constant $\Lambda$, the gravitational constant $G_{\text{N}}$, and the
speed of light $c$ are fundamental constants.

The overdot denotes differentiation with respect to $x^0=ct$.  The
Hubble function $H$ is defined by $H:=\dot{a}/a$.  For \textsc{flrw}
universes, the Hubble function is given by
$H = \tfrac{1}{3} \theta_\nul$, where $\theta_\nul$ is the background
value of the expansion scalar $\theta:=u^{\mu}{}_{;\mu}$, where
$u^\mu$ is the four-velocity $u^\mu:= c^{-1}U^\mu$, normalized to
unity, $u^\mu u_\mu=1$.  A semicolon denotes covariant differentiation
with respect to the background metric tensor $g_{\nul\mu\nu}$
(\ref{eq:metric-flrw}).  The equations~(\ref{FRW2}) and~(\ref{FRW2a})
represent the energy density conservation law and the particle number
conservation law, respectively.  The variable $w$ in Eq.\,(\ref{FRW2})
is an abbreviation for $p_{\nul}/\varepsilon_{\nul}$, and it does not
represent an equation of state.

The spatial parts of the background Riemann tensor
${^3\!}R^i_{\nul jkl}$, the Ricci tensor ${^3\!}R^i_{\nul j}$, and its
contraction ${^3\!}R_\nul$ are given by
\begin{subequations}
    \label{eq:glob-curve}
\begin{align}
   &  {^3\!}R^i_{\nul jkl}=\tilde{R}^i{}_{jkl}=K(\delta^i{}_k\tilde{g}_{jl}-
  \delta^i{}_l\tilde{g}_{jk}),  \label{eq:glob-curve-a} \\
  & {^3\!}R^i_{\nul j}=-\dfrac{2K}{a^2}\delta^i{}_j,
      \quad {^3\!}R_\nul=-\dfrac{6K}{a^2}, \label{eq:glob-curve-b}
\end{align}
\end{subequations}
where ${^3\!}R_\nul$ is the spatial curvature.  The value
of~$K$ determines the nature of the \textsc{flrw} universe. The
universe is open for $K=-1$, flat for $K=0$, and closed for $K=+1$.

\subsection{Linearized Equations}

\label{sec:lin-einstein-eq}

The system of linearized Einstein equations and conservation laws for
closed, flat, and open \textsc{flrw} universes is given by
\begin{subequations}
\label{subeq:basis}
\begin{align}
  &  H\dot{h}^k{}_k +\tfrac{1}{2}\,{^3\!}R_\een =
         -\kappa\varepsilon_\een,     \label{basis-1} \\
  & \dot{h}^k{}_{k\mid i}-\dot{h}^k{}_{i\mid k} =
         2\kappa\varepsilon_\nul(1 + w) u_{\een i}, \label{basis-2} \\
  & \ddot{h}^i{}_j+ 3H\dot{h}^i{}_j\, +\nonumber\\
  & \qquad \delta^i{}_j H\dot{h}^k{}_k
    +2\,{^3\!}R^i_{\een j} = -\kappa\delta^i{}_j(\varepsilon_\een-p_\een),
         \label{basis-3} \\
  & \dot{\varepsilon}_\een + 3H(\varepsilon_\een+p_\een)+
       \varepsilon_\nul(1+w)\theta_\een=0,
               \label{basis-4}  \\
  & \frac{1}{c}\frac{{\dif}}{{\dif} t}
    \left(\varepsilon_\nul(1+w) u^i_\een\right)\,- \nonumber \\
  & \qquad  g^{ik}_\nul p_{\een\mid k} +
   5H\varepsilon_\nul(1+w) u^i_\een=0,  \label{basis-5} \\
  & \dot{n}_\een+3Hn_\een+n_\nul\theta_\een = 0.  \label{basis-6}
\end{align}
\end{subequations}
In these equations, the perturbed metric tensor is given by
$h_{\mu\nu} := -g_{\een\mu\nu}$ and $h^{\mu\nu} := g^{\mu\nu}_{\een}$.
In light of the use of synchronous coordinates, it follows that
$h_{00}=0$ and $h_{0i}=0$.  Raising and lowering of indices is
performed by the spatial background metric tensor as defined in
expression (\ref{eq:metric-flrw}), namely,
$h^i{}_j=g^{ik}_\nul h_{kj}$ where
$g^{ik}_\nul=-\tilde{g}^{ik}/a^{2}$.  A vertical bar denotes covariant
differentiation with respect to the spatial background metric tensor
$g_{\nul ij}=-a^{2}\tilde{g}_{ij}$. In the context of \textsc{flrw}
universes, where $\Gamma^k_{\nul ij}=\tilde{\Gamma}^k{}_{ij}$, the
operations of taking the time derivative and the covariant derivative
are shown to commute. Equation~(\ref{basis-1}) represents the
linearized energy density constraint equation (linearized Friedmann
equation), while Eq.\,(\ref{basis-2}) corresponds to the linearized
momentum constraint equation. The linearized dynamical equations are
presented in Eqs.\,(\ref{basis-3}). Equations~(\ref{basis-4})
and~(\ref{basis-5}) represent the linearized energy density
conservation law and the linearized momentum conservation law,
respectively. Finally, the linearized particle number density
conservation law is given by Eq.\,(\ref{basis-6}).

Pressure perturbations, spatial curvature perturbations, and the
covariant divergence of the fluid velocity play pivotal roles in the
evolution of density perturbations. Therefore, these quantities must
be included in the perturbation theory.  The next three subsections
will derive expressions for these quantities.

\subsubsection{Pressure Perturbation}

The gauge-dependent quantity $p_{\een}$, which is associated with
pressure~(\ref{eq:equation-of-state-pressure}), is given by
\begin{subequations}
  \label{eq:p1}
  \begin{align}
    &  p_{\een}=p_{n}n_\een+p_\varepsilon\varepsilon_{\een}, \label{eq:p1-a} \\  
    &  p_{n}:=\left(\dfrac{\partial p}{\partial n}\right)_{\varepsilon}, \quad
      p_{\varepsilon}:=\left(\dfrac{\partial p}{\partial
      \varepsilon}\right)_{n}. \label{eq:p1-b}
\end{align}
\end{subequations}
The quantity $p^{\phys}_{\een}$ representing the pressure perturbation
will be derived in Sec.\,\ref{sec:thermodyn} on thermodynamics.

\subsubsection{Spatial Curvature Perturbation}
\label{subsec:spat-curv}

Lifshitz's expression for the perturbed connection coefficient
presented in Ref.\,\cite{c15}, Eq.\,(I.3) and Ref.\,\cite{c8},
Eq.\,(10.9.1) is a tensor, which is given by
\begin{equation}
    \Gamma^k_{\een ij}=-\tfrac{1}{2} g^{kl}_\nul
     (h_{li\mid j}+h_{lj\mid i}-h_{ij\mid l}). \label{con3pert}
\end{equation}
The contracted Palatini identity as defined in Ref.\,\cite{c15},
Eq.\,(I.5) and Ref.\,\cite{c8}, Eq.\,(10.9.2) is given by
\begin{equation}
   {^3\!}R_{\een ij} =
\Gamma^k_{\een ij\mid k}-\Gamma^k_{\een ik\mid j}.
    \label{palatini}
\end{equation}
By combining~(\ref{con3pert}) and~(\ref{palatini}), the following
expression for the perturbed spatial Ricci tensor can be derived (see
Ref.\,\cite{c8}, Eq.\,(10.9.3):
\begin{equation}
  \label{eq:Ricci-lower}
  {^3\!}R_{\een ij}=-\tfrac{1}{2} g^{kl}_\nul
     (h_{li\mid j\mid k}+h_{lj\mid i\mid k}-h_{ij\mid l\mid k}-h_{lk\mid i\mid j}).
\end{equation}
By raising the index $i$ and using~(\ref{eq:glob-curve-b}), one
arrives at
\begin{align}
  \label{eq:ricci-1}
    {^3\!}R^i_{\een j}:= &\, (g^{ik}\,{^3\!}R_{kj})_\een=
     g^{ik}_\nul\,{^3\!}R_{\een kj}+\tfrac{1}{3}\,{^3\!}R_\nul h^i{}_j\nonumber \\
     =&-\tfrac{1}{2}g^{il}_\nul(h^k{}_{l\mid j\mid k}+h^k{}_{j\mid l\mid k}-h^k{}_{k\mid l\mid j}) \nonumber\\
    & + \tfrac{1}{2}g^{kl}_\nul h^i{}_{j\mid k\mid l} +\tfrac{1}{3}\,{^3\!}R_\nul h^i{}_j.
\end{align}
By using
$g_\nul^{ij}h^k{}_{i\mid j\mid k}=g_\nul^{ij}h^k{}_{i\mid k\mid j}$,
one arrives at the contraction of the perturbed spatial Ricci tensor:
\begin{equation}
    {^3\!}R_\een  := {^3\!}R^k_{\een k} =
   g_\nul^{ij} (h^k{}_{k\mid i\mid j}-h^k{}_{i\mid k\mid j}) +
     \tfrac{1}{3}\,{^3\!}R_\nul h^k{}_k.   \label{driekrom}
\end{equation}
In Sec.\,\ref{sec:decomp-h-u} it will be demonstrated that the
expression~(\ref{driekrom}) represents the local perturbation to the
spatial curvature ${^3\!}R_{\nul}$ induced by local density
perturbations.  The quantity ${^3\!}R_\een$ will be used in
Sec.\,\ref{sec:scalar-pert} to eliminate the metric perturbations and
their spatial derivatives from the momentum constraint
equation~(\ref{basis-2}). It is incorporated into the perturbation
theory using Eq.\,(\ref{eq:sol-3R1}).

\subsubsection{Covariant Divergence of the Fluid Velocity}
\label{subsec:cov-div}

In the background \textsc{flrw} universe, the fluid four-velocity is
given by $u^\mu_\nul=\delta^\mu{}_0$.  Since $u^{\mu}u_{\mu} = 1$, it
follows that $u^{0}_{\een} = 0$, so that
$u^{\mu}_{\een} = (0, u^i_{\een})$.  The perturbation~$\theta_\een$ to
the expansion scalar $\theta:= u^\mu{}_{;\mu}$ is given by
\begin{equation}
  \theta_\een=
  \vartheta_\een-\tfrac{1}{2}\dot{h}^k{}_k, \quad
  \vartheta_\een:=u^k_{\een\mid k},
  \label{fes5}
\end{equation}
where $\vartheta_\een$ is the covariant divergence of the fluid
vector.  The quantity~$\theta_{\een}$ will be used in
Sec.\,\ref{sec:scalar-pert} to eliminate the time derivative of the
metric perturbation from the energy density constraint
equation~(\ref{basis-1}) in favor of~$\vartheta_{\een}$. The latter
quantity is incorporated into the perturbation theory using
Eq.\,(\ref{eq:sol-theta1}).

\section{Decomposition of Spatial Tensors and Vectors}
\label{sec:decomp-h-u}

The system of linearized Einstein equations and conservation
laws~(\ref{subeq:basis}) is not suitable for deriving the perturbation
theory because it contains solutions related to and unrelated to
density perturbations.  Therefore, our goal is to transform the
system~(\ref{subeq:basis}) into one that is exclusively linked to
density perturbations and excludes all other solutions.  To that end,
we will use two key theorems to break the linearized equations down
into three independent systems. These theorems are presented in the
next two subsections.

\subsection{Tensor Decomposition}

Refs.\,\cite{AIHPA_1974__21_4_319_0,SteWa,Stewart} demonstrate that any
rank-two symmetric spatial tensor ${\cal{T}}_{ij}$ can be decomposed
into three irreducible components in a covariant manner:
\begin{equation}
  {\cal{T}}^i{}_j={\cal{T}}^i_{\parallel\,j} +
  {\cal{T}}^i_{\perp\,j} + {\cal{T}}^i_{\ast\,j}. \label{eq:h123-a}
\end{equation}
The components have the following properties:
\begin{equation}
    {\cal{T}}^k_{\perp\,k}=0, \quad  {\cal{T}}^k_{\ast\,\, k}=0,
    \quad{\cal{T}}^k_{\ast\; i\mid k}=0, \quad
  {\cal{T}}^i_{\parallel\,j}=\phi\delta^i{}_j+\zeta^{\mid i}{}_{\mid j},  \label{eq:h123-b}
\end{equation}
where $\phi(t,\bm{x})$ and $\zeta(t,\bm{x})$ are two independent
potentials.  Consequently, the component $h^i_{\parallel\, j}$ of the
perturbed metric can be expressed in terms of the two independent
potentials as follows:
\begin{equation}
  h^i_{\parallel\, j} =
  \frac{2}{c^2}\bigl(\phi\delta^i{}_j+\zeta^{\mid i}{}_{\mid j}\bigr).
        \label{decomp-hij-par}
\end{equation}
The factor $2/c^{2}$ is included in view of the Newtonian limit.  By
substituting~(\ref{decomp-hij-par}) into~(\ref{driekrom}), the
following result is obtained:
\begin{align}
  \label{eq:driekrom-potentials}
  {^3\!}R_{\een\parallel}=\dfrac{2}{c^{2}}\Bigl[2\phi^{|i}{}_{|i}&+
  \zeta^{|k}{}_{|k}{}^{|i}{}_{|i}
    - \zeta^{|k|i}{}_{|k|i} \nonumber\\
   &+\tfrac{1}{3}\,{^3\!}R_{\nul}(3\phi+\zeta^{|k}{}_{|k})\Bigr],
\end{align}
and for the perturbed expansion~(\ref{fes5}) it is found
\begin{equation}
  \label{eq:expansion-potentials}
  \theta_{\een}=\vartheta_{\een}-\dfrac{1}{c^{2}}
  \bigl(3\dot{\phi}+\dot{\zeta}^{|k}{}_{|k}\bigr).
\end{equation}
The expression~(\ref{eq:driekrom-potentials}) reveals that two
potentials, $\phi(t,\bm{x})$ and $\zeta(t,\bm{x})$, are associated
with cosmological density perturbations in open and closed
\textsc{flrw} universes.  For a flat \textsc{flrw} universe, however,
density perturbations are characterized by one
potential,~$\phi(t,\bm{x})$. This is because, for scalar
perturbations, the quantity~$\theta_{\een}$ can be eliminated from
expressions~(\ref{e-n-gi}) and Eqs.\,(\ref{subeq:pertub-flrw}), as was
done in~(\ref{subeq:pertub-gi}) and~(\ref{subeq:pertub-gi-e-n}).

As will be demonstrated in Sec.\,\ref{sec:newt-limit}, in the
Newtonian limit, the potential~$\phi(t,\bm{x})$ becomes identical to
the Newtonian potential,~$\phi(\bm{x})$.

\subsection{Vector Decomposition}

Since $\bm{u}_\een$ is a vector in three-space, it can be decomposed
using the Helmholtz decomposition as follows:
\begin{equation}
   \bm{u}_{\een} = \bm{u}_{\een\parallel} +
  \bm{u}_{\een\perp}, \label{eq:decomp-u}
\end{equation}
where the components $\bm{u}_{\een\parallel}$ and $\bm{u}_{\een\perp}$
have the properties
\begin{equation} 
  \bm{\tilde{\nabla}}\bm{\cdot}\bm{u}_\een=
        \bm{\tilde{\nabla}}\bm{\cdot}\bm{u}_{\een\parallel}, \quad
  \bm{\tilde{\nabla}}
  \bm{\times}\bm{u}_\een=
     \bm{\tilde{\nabla}}\bm{\times}\bm{u}_{\een\perp}. \label{eq:nabla-u}
\end{equation}
In these expressions, the generalized vector differential operator,
denoted by $\bm{\tilde{\nabla}}$, is defined by
${\tilde{\nabla}_{i} v_{k}:= v_{k\mid i}}$.

Since $\bm{u}_{\een\parallel}$ is irrotational, it can be supplemented
with the gradient of an arbitrary function.  Given that the system of
equations~(\ref{subeq:basis}) is invariant under the gauge
transformation~(\ref{func}), where the gauge function
$\xi^{\mu}(t,\bm{x})$ is given by~(\ref{eq:synchronous}), it
follows~that
\begin{equation}
  \label{eq:gradient-to-ui}
  u^{\prime}_{\een\parallel i} = u_{\een\parallel i}+\psi_{|i},
\end{equation}
is also a solution to the system~(\ref{subeq:basis}).  As a result,
the gauge modes associated with $\bm{u}_{\een\parallel}$ are given by
\begin{equation}
  \label{eq:gauge-mode-upar}
  \hat{u}_{\een\parallel i}=\psi_{|i}, \quad
  \hat{u}^{i}_{\een\parallel}=g_{\nul}^{ik}\psi_{|k}=
  -\dfrac{1}{a^{2}}\tilde{g}^{ik}\psi_{|k}.
\end{equation}
This result will be used in Sec.\,\ref{sec:newt-limit} on the
Newtonian limit.

\section{Decomposition of the Linearized Equations}
\label{subsec:decomp-lin-eqs}

The system of equations~(\ref{subeq:basis}) is split into three
independent systems by using the decomposition~(\ref{eq:h123-a}) with
the properties~(\ref{eq:h123-b}) for $h_{ij}$ and ${^3\!}R_{ij}$, as
well as the decomposition~(\ref{eq:decomp-u}) with the
properties~(\ref{eq:nabla-u}) for $\bm{u}_{\een}$.  The solutions to
these systems are conventionally designated as tensor
perturbations~$\ast$, vector perturbations~$\perp$, and scalar
perturbations~$\parallel$.

In the following three subsections, it is demonstrated that scalar
perturbations are the only ones associated with density perturbations
$\varepsilon^\phys_\een$ and $n^\phys_\een$.

\subsection{Tensor Perturbations}

From the expressions ~(\ref{driekrom}) and~(\ref{eq:h123-b}), it
follows that ${^3\!}R_{\een\ast}=0$, and Eq.\,(\ref{basis-2}) yields
${\bm{u}_\een=\bm{0}}$, so that, with~(\ref{fes5}), $\theta_{\een}=0$.
Equation~(\ref{basis-1}) yields ${\varepsilon_{\een}=0}$.
From~(\ref{e-n-gi}), it follows that ${\varepsilon^{\phys}_{\een}=0}$.
Given that $\theta_{\een}=0$, Eqs.\,(\ref{FRW2a}) and~(\ref{basis-6})
are identical, so $n_{\een}=0$.  This, in turn, implies
with~(\ref{e-n-gi}), that $n^{\phys}_{\een}=0$.  Finally, the result
obtained from either~(\ref{basis-4}) or the contraction
of~(\ref{basis-3}) is that $p_{\een} = 0$.  Consequently, the
evolution equations for tensor perturbations are as follows:
\begin{equation}
  \label{eq:tensor-perturbations}
  \ddot{h}^i_{\ast\,j}+3H\dot{h}^i_{\ast\,j}+2\,{^3\!}R^i_{\een\ast j}=0.
\end{equation}
Given the form of these equations, tensor perturbations are typically
referred to as \emph{gravitational waves}.

\subsection{Vector Perturbations}

Expressions~(\ref{driekrom}) and~(\ref{eq:h123-b}), imply that, for
vector perturbations, we have
${^3\!}R_{\een\perp}=-h^{kj}_{\perp|k|j}$.  Raising the index~$i$ of
the momentum constraint equation~(\ref{basis-2}) with $g^{ij}_\nul$,
and subsequently taking the covariant derivative with respect to the
index~$j$, one finds, using that $\dot{g}^{ij}_\nul=-2Hg^{ij}_\nul$,
\begin{equation}
  \label{basis-2-raise}
  \dot{h}^{kj}_{\perp{\mid k\mid j}}+2Hh^{kj}_{\perp{\mid k\mid j}}=
  -2\kappa\varepsilon_\nul(1+w)u^j_{\een\perp\mid j}.
\end{equation}
The right-hand side of this equation vanishes, as can be seen
from~(\ref{eq:nabla-u}).  Consequently, the covariant divergence of
the momentum constraint equation~(\ref{basis-2-raise}) can only be
satisfied if ${^3\!}R_{\een\perp}=0$. The consequences are as follows.

Equation~(\ref{basis-1}) implies that ${\varepsilon_{\een}=0}$,
and~(\ref{fes5}) yields $\theta_{\een}=0$.  The latter identity shows
that Eq.\,(\ref{basis-6}) is identical to the background
equation~(\ref{FRW2a}), so ${n_{\een}=0}$. From~(\ref{e-n-gi}) we find
that $\varepsilon^{\phys}_{\een}=0$ and $n^{\phys}_{\een}=0$.
Finally, either~(\ref{basis-4}) or the contraction of~(\ref{basis-3})
implies that $p_{\een} = 0$.  Therefore, the system of equations for
vector perturbations is given by
\begin{subequations}
  \label{subeq:vector-perturbations}
  \begin{align}
   & \!\dot{h}^k_{\perp i\mid k}+2\kappa\varepsilon_\nul(1+w)u_{\een\perp i}=0, \\
   & \!\ddot{h}^i_{\perp j}+3H\dot{h}^i_{\perp j}+2\,{^3\!}R^i_{\een\perp j}=0, \\
   & \!\frac{1}{c}\frac{{\text{d}}}{{\text{d}} t}
    \Bigl(\varepsilon_\nul(1+w) u^i_{\een\perp}\Bigr)+
    5H\varepsilon_\nul(1+w) u^i_{\een\perp}=0.
  \end{align}
\end{subequations}
According to~(\ref{eq:nabla-u}), vector perturbations are also
referred to as \emph{rotational perturbations}.

\subsection{Scalar Perturbations}

In this particular instance one has ${^3\!}R_{\een\parallel}\neq0$, as
follows from expression~(\ref{eq:driekrom-potentials}).  By taking the
covariant derivative of the momentum constraint
equation~(\ref{basis-2}) with respect to the index $j$ and
subsequently substituting the expression~(\ref{decomp-hij-par}), the
following result is obtained:
\begin{equation}
  2\dot{\phi}_{\mid i\mid j}+\dot{\zeta}^{\mid k}{}_{\mid k\mid i\mid j}-
         \dot{\zeta}^{\mid k}{}_{\mid i\mid k\mid j} =
         \kappa c^2\varepsilon_\nul(1 + w) u_{\een i\mid j}.
    \label{eq:feiko1}
\end{equation}
By interchanging the indices~$i$ and~$j$ and subtracting the result
from Eqs.\,(\ref{eq:feiko1}) one finds
\begin{equation}
  \dot{\zeta}^{\mid k}{}_{\mid i\mid k\mid j}-
  \dot{\zeta}^{\mid k}{}_{\mid j\mid k\mid i} =
   -\kappa c^2\varepsilon_\nul(1 + w)(u_{\een i\mid j}-u_{\een j\mid i}),
\label{dR0i2-rot}
\end{equation}
where it is used that
$\dot{\phi}_{\mid i\mid j}=\dot{\phi}_{\mid j\mid i}$ and
$\dot{\zeta}^{\mid k}{}_{\mid k\mid i\mid j}=\dot{\zeta}^{\mid
  k}{}_{\mid k\mid j\mid i}$.  Rearranging the second-order covariant
derivatives converts Eq.\,(\ref{dR0i2-rot}) into a form with only
commutators on the left-hand side:
\begin{align}
    & (\dot{\zeta}^{\mid k}{}_{\mid i\mid k\mid j}  - \dot{\zeta}^{\mid k}{}_{\mid i\mid j\mid k})
      -(\dot{\zeta}^{\mid k}{}_{\mid j\mid k\mid i}-\dot{\zeta}^{\mid k}{}_{\mid j\mid i\mid k})\,+   \nonumber \\
    & \qquad   (\dot{\zeta}^{\mid k}{}_{\mid i\mid j}-\dot{\zeta}^{\mid k}{}_{\mid j\mid i})_{\mid k}\,= \nonumber\\
    & \qquad  -\kappa c^2\varepsilon_\nul(1 + w)
      (u_{\een i\mid j}-u_{\een j\mid i}).
     \label{eq:verwissel}
\end{align}
The commutators of second-order covariant derivatives (see
Ref.\,\cite{c8}, Sec.\,6.5) are given~by
\begin{subequations}
  \label{eq:commu-Riemann}
  \begin{align}
 & A^i{}_{j\mid p\mid q}-A^i{}_{j\mid q\mid p} =
      A^i{}_k \,{^3\!}R^k_{\nul jpq} - A^k{}_j\,{^3\!}R^i_{\nul kpq}, \\
  &  B^i{}_{\mid p\mid q}-B^i{}_{\mid q\mid p} = B^k \,{^3\!}R^i_{\nul kpq}.
  \end{align}
\end{subequations}
Upon substituting the background Riemann
tensor~(\ref{eq:glob-curve-a}), the left-hand side of
Eq.\,(\ref{eq:verwissel}) vanishes identically.  This implies that
$\bm{\tilde{\nabla}}\bm{\times}\bm{u}_\een=\bm{0}$, so that only the
component $\bm{u}_{\een\parallel}$ remains. Since
$\bm{\tilde{\nabla}}\bm{\cdot}\bm{u}_{\een\parallel}\neq0$ implies
that $\varepsilon^\phys_\een\neq0$ and $n^\phys_\een\neq0$, it can
thus be concluded that only scalar perturbations are coupled to
density perturbations.

\section{Evolution Equations for Scalar Perturbations}
\label{sec:scalar-pert}

As demonstrated in the preceding section, the evolution of
$\varepsilon^\phys_\een$ and $n^\phys_\een$ is determined by the
background equations~(\ref{subeq:einstein-flrw}) and the equations
that describe the evolution of scalar perturbations.  In this section,
the findings from the previous section are used to reformulate the
system~(\ref{subeq:basis}) into a new system of equations that
exclusively describes the evolution of scalar perturbations.  Given
the subsequent focus on density perturbations, the
subscript~$\parallel$ is omitted.  The evolution equations for scalar
perturbations are as follows:
\begin{subequations}
\label{subeq:pertub-flrw}
\begin{align}
 &  2H(\theta_\een-\vartheta_\een) =
       \tfrac{1}{2}\,{^3\!}R_\een + \kappa\varepsilon_\een, \label{con-sp-1} \\
  & \!\! {^3\!}\dot{R}_\een = -2H\,{^3\!}R_\een  \nonumber \\
  & \qquad   \;\,   +2\kappa \varepsilon_\nul(1+w)\vartheta_\een
         -\tfrac{2}{3}\,{^3\!}R_\nul (\theta_\een-\vartheta_\een), \label{FRW6} \\
 &  \dot{\varepsilon}_\een = - 3H(\varepsilon_\een + p_\een)-
         \varepsilon_\nul(1 + w)\theta_\een,  \label{FRW4} \\
  &  \dot{\vartheta}_\een = -H(2-3\beta^2)\vartheta_\een
 -\frac{1}{\varepsilon_\nul(1+w)}\dfrac{\tilde{\nabla}^2p_\een}{a^2}, \label{FRW5}\\
 &  \dot{n}_\een = - 3H n_\een - n_\nul\theta_\een. \label{FRW4a}
\end{align} 
\end{subequations}
These equations describe the evolution of the five gauge-dependent
quantities $\theta_{\een}$, $\vartheta_{\een}$, ${^3\!}R_\een$,
$\varepsilon_{\een}$, and $n_{\een}$. The associated gauge modes are
given by~(\ref{subeq:gauge-dep}).  The potentials~$\phi(t,\bm{x})$
and~$\zeta^{|i}{}_{|j}(t,\bm{x})$ are encapsulated in the
quantities~${^3\!}R_\een$ (\ref{eq:driekrom-potentials})
and~$\theta_{\een}$ (\ref{eq:expansion-potentials}).  This greatly
simplifies the linearized Einstein equations and conservation
laws~(\ref{subeq:basis}) when studying density perturbations.
Additionally, the system~(\ref{subeq:pertub-flrw}) allows us to derive
the Newtonian limit in Sec.\,\ref{sec:newt-limit}.

The symbol $\tilde{\nabla}^2$ denotes the Laplace-Beltrami operator
with respect to the metric $\tilde{g}_{ij}(\bm{x})$ of
three-dimensional subspaces of constant time:
\begin{equation}
  \label{eq:lap-bel-op-flrw}
  \tilde{\nabla}^2p_{\een}:=\tilde{g}^{ij} p_{\een\mid i\mid j}, \quad
  g^{ij}_{\nul}p_{\een|i|j}=
       -\dfrac{\tilde{\nabla}^2p_{\een}}{a^{2}}.
\end{equation}
The parameter~$\beta$ is defined as follows:
\begin{equation}
  \label{eq:parameter-beta}
  \beta^{2}:=\dfrac{\dot{p}_{\nul}}{\dot{\varepsilon}_{\nul}}.
\end{equation}
Using the expression
$\dot{p}_\nul=p_n\dot{n}_\nul+p_\varepsilon\dot{\varepsilon}_\nul$ and
eliminating the time derivatives of $\varepsilon_\nul$ and $n_\nul$
with the aid of the conservation laws~(\ref{FRW2}) and~(\ref{FRW2a}),
the following result is obtained:
\begin{equation}
  \label{eq:beta-matter}
  \beta^2=p_\varepsilon+\dfrac{n_\nul p_n}{\varepsilon_\nul(1+w)},
\end{equation}
where $p_\varepsilon$ and $p_n$ are given by~(\ref{eq:p1-b}).

We will now derive Eqs.\,(\ref{subeq:pertub-flrw}).  To begin, the
term $\dot{h}^k{}_k$ is eliminated from Eq.\,(\ref{basis-1}) using the
expression~(\ref{fes5}). This results in the algebraic energy density
constraint equation~(\ref{con-sp-1}).

Given that only $\bm{u}_{\een\parallel}$ is associated with scalar
perturbations, it is possible to simplify the momentum constraint
equation~(\ref{basis-2}) and the momentum conservation
law~(\ref{basis-5}) by using the covariant divergence
${\vartheta_{\een}:=\bm{\tilde{\nabla}}\bm{\cdot}\bm{u}_{\een\parallel}}$,
(\ref{fes5}), as an alternative to $\bm{u}_{\een\parallel}$.  By
multiplying both sides of Eq.\,(\ref{basis-2}) by $g^{ij}_\nul$ and
taking the covariant derivative with respect to the index $j$, one
finds
\begin{equation}
  g_\nul^{ij} (\dot{h}^k{}_{k\mid i\mid j}-\dot{h}^k{}_{i\mid k\mid j}) =
    2\kappa\varepsilon_\nul(1+w) \vartheta_\een.  \label{dR0i4}
\end{equation}
The left-hand side of Eq.\,(\ref{dR0i4}) will appear as a component of
the time derivative of the local spatial curvature perturbation,
${^3\!}R_\een$.  Indeed, differentiating~(\ref{driekrom}) with respect
to time, and using the relation $\dot{g}^{ij}_\nul=-2Hg^{ij}_\nul$
and~(\ref{FRW3a}), yields the following result:
\begin{equation}
  {^3\!}\dot{R}_\een = -2H\,{^3\!}R_\een +
   g_\nul^{ij} (\dot{h}^k{}_{k\mid i\mid j}-\dot{h}^k{}_{i\mid k\mid j})+
     \tfrac{1}{3}\,{^3\!}R_\nul \dot{h}^k{}_k.
   \label{dR0i6}
\end{equation}
By combining~(\ref{dR0i4}) and~(\ref{dR0i6}) and using~(\ref{fes5}) to
eliminate $\dot{h}^k{}_k$, one obtains Eq.\,(\ref{FRW6}).
Consequently, the $G^0_{\een i}$ momentum constraint
equation~(\ref{basis-2}) has been reformulated as a first-order
ordinary differential equation~(\ref{FRW6}) for the perturbed spatial
Ricci scalar~(\ref{eq:driekrom-potentials}) expressed in the
potentials.

The momentum conservation law~(\ref{basis-5}) is reformulated by first
performing the differentiation with respect to time.  Then, we use
Eq.\,(\ref{FRW2}) to eliminate $\dot{\varepsilon}_{\nul}$.  Next,
dividing the result by $\varepsilon_{\nul}(1+w)$ yields
\begin{equation}
  \label{eq:mom-w-punt}
 \dot{u}^{i}_{\een}+H(2-3w)u^{i}_{\een}+
  \dfrac{\dot{w}}{1+w}u^{i}_{\een}-
  \dfrac{g^{ik}_\nul p_{\een\mid k}}{\varepsilon_{\nul}(1+w)}=0.
\end{equation}
This equation can be simplified by eliminating the time derivative
of~$w$.  From the definitions $w:= p_\nul/\varepsilon_\nul$ and
$\beta^2:=\dot{p}_\nul/\dot{\varepsilon}_\nul$ and the energy
conservation law~(\ref{FRW2}) one obtains
\begin{equation}
  \label{eq:time-w}
  \dot{w}=3H(1+w)(w-\beta^2).
\end{equation}
Upon substituting this expression into Eq.\,(\ref{eq:mom-w-punt}), the
momentum conservation law becomes
\begin{equation}
  \label{eq:mom2}
  \dot{u}^{i}_\een + H(2-3\beta^2)u^{i}_\een
   -\frac{g^{ik}_{\nul} p_{\een|k}}{\varepsilon_\nul(1+w)}=0.
\end{equation}
Taking the covariant divergence of~(\ref{eq:mom2}) with respect to the
spatial background metric tensor~(\ref{eq:metric-flrw}), and using
(\ref{fes5}) and~(\ref{eq:lap-bel-op-flrw}) yields Eq.\,(\ref{FRW5}).

As demonstrated in Appendix\;\ref{app:consistency-check}, the
dynamical equation~(\ref{basis-3}) is redundant since the
system~(\ref{subeq:pertub-flrw}) comprises the constraint equations
and conservation laws.  Thus, the derivation of
Eqs.\,(\ref{subeq:pertub-flrw}) has been completed.

\section{Newtonian Limit}
\label{sec:newt-limit}

As demonstrated in Sec.\,\ref{sec:overview-gauge-problem}, the
gauge-invariant quantities, $\varepsilon^\phys_\een$ and
$n^\phys_\een$, which are intended to describe density perturbations,
can only be defined in one way.  This section proves that the
quantities $\varepsilon^\phys_\een$ and $n^\phys_\een$ represent
perturbations to energy and particle number densities, respectively.

In Newtonian gravity, space is Euclidean. Consequently, the Newtonian
limit can only be realized in a flat \textsc{flrw} universe. In this
case, the Newtonian limit is reached when gravity is weak and the
velocities of the particles are negligible compared to the speed of
light.  These conditions are met in a linear perturbation theory and
when pressure approaches zero.

In a flat \textsc{flrw} universe where ${^3\!}R_\nul=0$, covariant
derivatives are equal to ordinary derivatives. This simplifies the
expression~(\ref{eq:driekrom-potentials}) to
\begin{equation}\label{RnabEE-0}
   {^3\!}R_{\een} =\dfrac{4}{c^2}\phi^{\mid k}{}_{\mid k}=
       -\dfrac{4}{c^2}\dfrac{\nabla^2\phi}{a^2}.
\end{equation}
The symbol $\nabla^{2}$ represents the conventional Laplace
operator. By substituting the expression~(\ref{RnabEE-0}) into the
perturbation equations~(\ref{subeq:pertub-flrw}) and using that
$H := \dot{a}/a$, we obtain the result:
\begin{subequations}
\label{subeq:pertub-gi-flat}
\begin{alignat}{3}
     -H\vartheta_\een
       & =-\dfrac{1}{c^2}\dfrac{\nabla^2\phi}{a^2}+ \tfrac{1}{2}\kappa
        \varepsilon^\phys_\een,  \quad  \kappa=8\pi G_{\text{N}}/c^{4},  \label{con-sp-1-flat} \\
      \dfrac{1}{c^2}\dfrac{\nabla^2\dot{\phi}}{a^2} & = -\tfrac{1}{2}\kappa
         \varepsilon_\nul(1 + w)\vartheta_\een, \quad  w:=p_{\nul}/\varepsilon_{\nul}, \label{FRW6gi-flat} \\
     \dot{\varepsilon}_\een & =-3H(\varepsilon_\een + p_\een)-
         \varepsilon_\nul(1 + w)\theta_\een,   \label{FRW4gi-flat} \\ 
      \dot{\vartheta}_\een &  = -H(2-3\beta^2)\vartheta_\een-
      \frac{1}{\varepsilon_\nul(1+w)}\dfrac{\nabla^2p_\een}{a^2},  \label{FRW5gi-flat} \\
    \dot{n}_\een & = - 3H n_\een -
         n_\nul\theta_\een, \quad \beta^{2}:=\dot{p}_{\nul}/\dot{\varepsilon}_{\nul}. \label{FRW4agi-flat}
\end{alignat}
\end{subequations}
To derive the $G^0_{\een0}$ constraint equation~(\ref{con-sp-1-flat})
from Eq.\,(\ref{con-sp-1}), the function $\varepsilon_{\een}$ must
first be eliminated using the expression~(\ref{e-n-gi}).  The result
is
\begin{equation}
  \label{eq:tussen-resultaat}
  2H(\theta_{\een}-\vartheta_{\een})=
  -\dfrac{2}{c^2}\dfrac{\nabla^2\phi}{a^2}+
  \kappa\left(\varepsilon^\phys_\een+
    \dfrac{\dot{\varepsilon}_{\nul}}{\dot{\theta}_{\nul}}\theta_{\een}   \right).
\end{equation}
Setting ${^3\!}R_\nul=0$ in the Friedmann equation~(\ref{FRW3}) and
differentiating the result with respect to time, and using the
relation $H=\tfrac{1}{3}\theta_\nul$, we obtain
$\dot{\varepsilon}_\nul/\dot{\theta}_\nul=2H/\kappa$.  Substituting
this expression into Eq.\,(\ref{eq:tussen-resultaat}) yields the
desired result, Eq.\,(\ref{con-sp-1-flat}).  The Newtonian limit can
now be derived using the system~(\ref{subeq:pertub-gi-flat}).

In a universe filled with a nonrelativistic fluid, we have $w\ll1$ and
$\beta^{2}\ll1$.  Neglecting the quantities $w$ and $\beta^{2}$ with
respect to the constants of order one in the momentum conservation
law~(\ref{eq:mom2}), yields
\begin{equation}
  \label{eq:mom2-bw-nul}
    \dot{u}^{i}_\een + 2Hu^{i}_\een
   -\frac{g^{ik}_{\nul} p_{\een|k}}{\varepsilon_\nul}=0, \quad
  \dot{u}_{\een j} - \dfrac{p_{\een|j}}{\varepsilon_\nul}=0.
\end{equation}
The second equation is derived by lowering the index $i$ in the first
equation using the background metric $g_{\nul ij}$ and the relation
$\dot{g}_{\nul ij} = 2Hg_{\nul ij}$.  Accordingly, the evolution of
density perturbations is highly dependent upon the presence of
pressure gradients, represented by~$p_{\een|j}$, where $p_{\een}$ is
given by~(\ref{eq:p1}).  In a \emph{pressureless} fluid, i.e., when
$w=0$ and $\beta=0$, however, the result is
\begin{equation}
  \label{eq:ui-par-p0}
   \dot{u}^{i}_{\een}+2Hu^{i}_{\een}=0, \quad \dot{u}_{\een j}=0.
\end{equation}
The second equation has solutions that are only functions of the
spatial coordinates.  Since these equations are part of the
system~(\ref{subeq:basis}) that is invariant under the gauge
transformation~(\ref{func}) with $\xi^{\mu}(t,\bm{x})$ given
by~(\ref{eq:synchronous}), it can be inferred that the gauge modes
given by~(\ref{eq:gauge-mode-upar}), are the only solutions to
equations~(\ref{eq:ui-par-p0}).  Therefore, it can be concluded that
\begin{equation}
  \label{eq:non-rel-lim}
  p\rightarrow0 \quad
    \Rightarrow \quad u_{\een i}^{\phys}(t,\bm{x}) \rightarrow 0.
\end{equation}
In other words, in the absence of pressure, local pressure
gradients~$p_{\een|i}$ are nonexistent, so local fluid flows are
impossible. Thus, in the Newtonian limit, density perturbations are
independent of time, that is, they are \emph{static}.  The remaining
part of this section examines the consequences of the
limit~(\ref{eq:non-rel-lim}).

Since the components~(\ref{eq:non-rel-lim}) of the fluid velocity are
zero, only the gauge modes $\hat{u}_{\een i}=\psi_{|i}$ remain.  We
can set these gauge modes to zero without losing any physical
information.  Substituting $\psi_{|i}=0$ into the
expressions~(\ref{eq:synchronous}) yields $\xi^{0}=C$ and
$\xi^{i}=\chi^{i}$, implying that the relativistic gauge
transformation~(\ref{func}) reduces to the gauge transformation of
Newtonian gravity:
\begin{equation}
  \label{eq:gauge-trans-newt}
  t \rightarrow t^\prime=t-C, \quad   x^i \rightarrow x^{\prime i}=x^i - \chi^i(\bm{x}).
\end{equation}
Thus, we arrive at the well-known fact that in Newtonian gravity, we
can freely select the spatial coordinates and shift the time
coordinate at will, and that space and time are independent of each
other.

Since both $\hat{u}_{\een i}=0$ and $u_{\een i}^{\phys}=0$
we find from~(\ref{fes5}) that $\vartheta_\een=0$. By substituting the
latter into the system~(\ref{subeq:pertub-gi-flat}) we obtain
\begin{subequations}
\label{subeq:pertub-gi-flat-newt}
\begin{align}
   \nabla^2\phi(t,\bm{x}) &= \dfrac{4\pi G_{\text{N}}}{c^2} a^2(t)\varepsilon^\phys_\een(t,\bm{x}),
         \label{con-sp-1-flat-newt} \\
   \nabla^2\dot{\phi}(t,\bm{x}) &= 0, \label{FRW6gi-flat-newt} \\
  \dot{\varepsilon}_\een &= - 3H\varepsilon_\een-\varepsilon_\nul\theta_\een,
               \label{FRW4gi-flat-newt} \\
   \dot{n}_\een &= - 3H n_\een-n_\nul\theta_\een. \label{FRW4agi-flat-newt}
\end{align}
\end{subequations}
The constraint equations~(\ref{con-sp-1-flat-newt})
and~(\ref{FRW6gi-flat-newt}) imply
\begin{equation}
  \label{eq:at-at-at0-et0}
  a^2(t)\varepsilon^\phys_\een(t,\bm{x})=
      a^2(t_{0})\varepsilon^\phys_\een(t_{0},\bm{x}).
\end{equation}
This equation applies at any given time $t_{0}$.  From the Friedmann
equation~(\ref{FRW3}), it follows that, for ${^3\!}R_\nul=0$, one may
take $a(t_{0})=1$, so that
\begin{equation}
  \label{eq:e-phys-static}
  a^2(t)\varepsilon^\phys_\een(t,\bm{x})=\varepsilon^\phys_\een(\bm{x}).
\end{equation}
This expression is consistent with the limit~(\ref{eq:non-rel-lim})
that results in static density perturbations.
Combining~(\ref{con-sp-1-flat-newt}), (\ref{FRW6gi-flat-newt}),
and~(\ref{eq:e-phys-static}) yields
\begin{equation}
  \label{eq:poisson}
  \nabla^2\phi(\bm{x})=4\pi G_{\text{N}} \dfrac{\varepsilon^\phys_\een(\bm{x})}{c^2},
\end{equation}
which is the Poisson equation of Newtonian gravity.

Finally, the meaning of $n^\phys_\een$ is established.  For a universe
filled with a pressureless fluid, the background
equations~(\ref{subeq:einstein-flrw}) are given~by
\begin{equation}
  \label{eq:einstein-flrw-nonrel}
  3H^2 =\kappa\varepsilon_\nul+\Lambda, \quad
    \dot{\varepsilon}_\nul  =-3H\varepsilon_\nul,\quad
    \dot{n}_\nul =-3H n_\nul.
\end{equation}
Consequently, the conservation laws imply that
$\varepsilon_{\nul}=n_{\nul} mc^2$.  As can be verified by
substitution, Eqs.\,(\ref{FRW4gi-flat-newt})--(\ref{FRW4agi-flat-newt})
yield solely gauge modes as solutions: specifically,
$\hat{\varepsilon}_\een=C\dot{\varepsilon}_\nul$,
$\hat{n}_{\een}=C\dot{n}_{\nul}$, and
$\hat{\theta}_{\een}=C\dot{\theta}_{\nul}$, where $C$ is an
infinitesimal constant of the gauge
transformation~(\ref{eq:gauge-trans-newt}) of Newtonian gravity. This,
combined with $\varepsilon_{\nul}=n_{\nul} mc^2$ yields
$\varepsilon_{\een}=n_{\een} mc^2$.  Combining the latter two
relations with the definitions~(\ref{e-n-gi}) one finds
\begin{equation}
  \label{eq:newt-ngi}
    \varepsilon^\phys_\een(\bm{x})=n^\phys_\een(\bm{x})mc^2.
\end{equation}
This is the well-known special relativistic relation between rest mass
and energy.

The equations~(\ref{FRW4gi-flat-newt}) and~(\ref{FRW4agi-flat-newt})
represent the residual form of the conservation
laws~(\ref{FRW4gi-flat}) and~(\ref{FRW4agi-flat}) in the Newtonian
limit.  Since Eqs.\,(\ref{FRW4gi-flat-newt})
and~(\ref{FRW4agi-flat-newt}) only have gauge modes as solutions, they
clearly have no physical significance.  These equations are decoupled
from the physical constraint equations~(\ref{con-sp-1-flat-newt})
and~(\ref{FRW6gi-flat-newt}) and thus do not form part of Newtonian
gravity. Therefore, the potential $\zeta$ which appears in
$\theta_{\een}$~(\ref{eq:expansion-potentials}) is no longer
relevant. This is consistent with the comment just below
(\ref{eq:expansion-potentials}).  Thus, in the Newtonian limit, only
the Poisson equation~(\ref{eq:poisson}) and the special relativistic
relation~(\ref{eq:newt-ngi}) remain.

We now come to a final conclusion.  The analysis above demonstrates
that, in the Newtonian limit, only the Poisson
equation~(\ref{eq:poisson}) and the relation~(\ref{eq:newt-ngi})
remain.  Therefore, $\varepsilon_\een^\phys$ and $n_\een^\phys$
(\ref{e-n-gi}) represent the perturbations in energy and particle
number density, respectively, in Newtonian gravity.  Consequently,
they also represent perturbations in energy and particle number
density in general relativity.  Finally, we note that, since
$\theta^{\phys}_{\een}=0$ (\ref{theta-gi}), density perturbations do
not affect the expansion of the universe.

In the next subsections, relative density perturbations in the
Newtonian limit are discussed, and the perturbation theory based on
Newtonian gravity is examined in the context of relativity theory.

\subsection{Relative Density Perturbation in the Newtonian Limit}

The expression~(\ref{eq:e-phys-static}) implies that
\begin{equation}
  \label{eq:e-phys-a2}
  \varepsilon^\phys_\een(t,\bm{x}) =
  \dfrac{\varepsilon^\phys_\een(\bm{x})}{a^{2}(t)}.
\end{equation}
Since Eq.\,(\ref{FRW2}) with $w=0$ implies
$\varepsilon_{\nul}\propto a^{-3}$, it follows that the relative
density perturbation~$\delta_{\varepsilon}$ evolves according~to
\begin{equation}
  \label{eq:fluct-p-is-nul}
  \delta_{\varepsilon}(t,\bm{x}):=\dfrac{\varepsilon^\phys_\een(t,\bm{x})}
      {\varepsilon_{\nul}(t)} \propto \varepsilon^\phys_\een(\bm{x}) a(t).
\end{equation}
Thus, due to the expansion of the universe, the relative density
perturbation $\delta_{\varepsilon}$ for a \emph{static} density
perturbation $\varepsilon^\phys_\een(\bm{x})$ increases proportionally
to the scale factor $a(t)$ of the universe. In other words, a static
density perturbation becomes relatively denser when viewed against the
decreasing energy density~$\varepsilon_{\nul}$ in the background
universe.

\subsection{Newtonian Gravity}

As demonstrated in Appendix\;\ref{sec:stand-theory}, the relative
perturbation $\delta:=\varepsilon_{\een}/\varepsilon_{\nul}$ is
gauge-dependent.  The gauge mode associated with~$\delta$ is given
by~(\ref{eq:delta-e--gauge}). For $w\ll1$, it is a solution to the
homogeneous part of Eq.\,(\ref{eq:alg-mat-dom-delta}).  This gauge
mode does not disappear from the solution because $H\neq0$ in a
nonempty universe. Consequently, the quantity~$\delta$ depends on the
choice of coordinates and therefore has no physical meaning.  As can
be seen from Eqs.\,(\ref{eq:einstein-flrw-nonrel}), $H\neq0$ even in
the Newtonian limit.  Therefore, $\delta$ must also depend on the
choice of coordinates in that limit and consequently has no physical
significance.  This is consistent with the fact that, in the Newtonian
limit, the relativistic gauge transformation~(\ref{func}) converges to
the Newtonian gauge transformation~(\ref{eq:gauge-trans-newt}).  Given
that the system~(\ref{subeq:algemene-vergelijkingen}) reduces to the
system~(\ref{subeq:algemene-vergelijkingen-mat-dom}) for a
nonrelativistic fluid, it can be concluded that the
system~(\ref{subeq:algemene-vergelijkingen-mat-dom}) is inadequate for
examining the evolution of density perturbations.

The existing literature derives Eq.\,(\ref{eq:standaard-after-dec}),
from Newtonian gravity under two conditions: the expansion of the
universe must be considered and the equation of state must be
nonrelativistic, i.e., $w:=p_{\nul}/\varepsilon_{\nul}\ll1$ and
$\beta^{2}\ll1$.  See, for example, Ref.\,\cite{Mukhanov-2005},
Sec.\,6.2, Ref.\,\cite{c8}, Sec.\,15.9, and
Ref.\,\cite{10.1093/mnras/117.1.104}, Eq.\,(3.17).  However, the gauge
transformation~(\ref{eq:gauge-trans-newt}) shows that~$\delta$ depends
on the choice of coordinates in Newtonian gravity. Therefore, it also
lacks physical significance in Newtonian gravity.  Furthermore, a
comparison of the conventional equation~(\ref{eq:standaard-after-dec})
and the exact equation~(\ref{eq:alg-mat-dom-delta}) reveals that
equation~(\ref{eq:standaard-after-dec}) is incomplete.  Consequently,
the conventional evolution equation~(\ref{eq:standaard-after-dec}) ---
whether derived from general relativity or Newtonian gravity modified
to account for the expansion of the universe --- also has
gauge-dependent solutions.  These findings indicate that Newtonian
gravity is inadequate in accurately describing the evolution of
cosmological density perturbations.  Therefore, there exists no
distinction between sub- and super-horizon density perturbations.

\section{Relativistic Cosmological Perturbation Theory}
\label{sec:flat-pert}

As demonstrated in Sec.\,\ref{subsec:decomp-lin-eqs}, the evolution
equations for scalar perturbations~(\ref{subeq:pertub-flrw}) when
combined with the background equations~(\ref{subeq:einstein-flrw})
describe the evolution of energy density and particle number density
perturbations.  In Appendix\;\ref{app:computer-algebra}, the
derivation of evolution equations for their corresponding relative
density perturbations defined by
\begin{equation}
  \label{eq:contrast}
  \delta_\varepsilon(t,\bm{x}) :=
  \dfrac{\varepsilon^\phys_\een(t,\bm{x})}{\varepsilon_\nul(t)}, \quad
  \delta_n(t,\bm{x}) :=
  \dfrac{n^\phys_\een(t,\bm{x})}{n_\nul(t)},
\end{equation}
will be presented.  The result is a system of evolution equations for
the relative density perturbations~$\delta_{\varepsilon}$
and~$\delta_{n}$, in closed, flat, and open \textsc{flrw} universes,
given by
\begin{subequations}
\label{subeq:final}
\begin{align}
   \ddot{\delta}_\varepsilon + b_1 \dot{\delta}_\varepsilon +
      b_2 \delta_\varepsilon &=
      b_3 \left(\delta_n - \frac{\delta_\varepsilon}{1+w}\right),
              \label{sec-ord}  \\
  \frac{1}{c}\frac{\dif}{\dif t}
      \left(\delta_n - \frac{\delta_\varepsilon}{1 + w}\right) & =
     \frac{3Hn_\nul p_n}{\varepsilon_\nul(1 + w)}
     \left(\delta_n - \frac{\delta_\varepsilon}{1 + w}\right).
                 \label{fir-ord}
\end{align}
\end{subequations}
The background equations~(\ref{subeq:einstein-flrw}) and an equation
of state~(\ref{eq:equation-of-state-pressure}) for the pressure
determine the coefficients $b_1$, $b_2$, and $b_3$. These coefficients
are given by
\begin{subequations}
\label{subeq:coeff-contrast}
\begin{align}
  b_1  = &\, \dfrac{\kappa\varepsilon_\nul(1+w)}{H}
  -2\dfrac{\dot{\beta}}{\beta}-H(2+6w+3\beta^2) \nonumber \\
  & + {^3\!}R_\nul\left(\dfrac{1}{3H}+
    \dfrac{2H(1+3\beta^2)}{{^3\!}R_\nul+3\kappa\varepsilon_\nul(1+w)}\right),
       \label{eq:b1} \\
  b_2 = & -\tfrac{1}{2}\kappa\varepsilon_\nul(1+w)(1+3w) \nonumber \\
  &+ H^2\left(1-3w+6\beta^2(2+3w)\right) \nonumber\\
   &+6H\dfrac{\dot{\beta}}{\beta}\left(w+\dfrac{\kappa\varepsilon_\nul(1+w)}
   {{^3\!}R_\nul+3\kappa\varepsilon_\nul(1+w)}\right)  \nonumber \\
   & - {^3\!}R_\nul\left(\tfrac{1}{2}w+
\dfrac{H^2(1+6w)(1+3\beta^2)}{{^3\!}R_\nul+3\kappa\varepsilon_\nul(1+w)}\right) \nonumber\\
&  -\beta^2\left(\frac{\tilde{\nabla}^2}{a^2}-\tfrac{1}{2}\,{^3\!}R_\nul
\right), \label{eq:b2} \\
  b_3  =&\,
\Biggl[\dfrac{-18H^2}{{^3\!}R_\nul+3\kappa\varepsilon_\nul(1+w)}
  \Biggl(\varepsilon_\nul p_{\varepsilon n}(1+w)
   +\dfrac{2p_n}{3H}\dfrac{\dot{\beta}}{\beta} \nonumber\\
   & +p_n(p_\varepsilon-\beta^2)+n_\nul p_{nn}\Biggr)+
     p_n\Biggr] \nonumber \\
   & \times  \dfrac{n_\nul}{\varepsilon_\nul}
    \left(\frac{\tilde{\nabla}^2}{a^2}-\tfrac{1}{2}\,{^3\!}R_\nul\right).
\label{eq:b3}
\end{align}
\end{subequations}
In these expressions the partial derivatives of the pressure,
$p_\varepsilon$ and $p_n$, are defined by~(\ref{eq:p1-b}). The
second-order partial derivatives are defined by
$p_{nn}:={\partial^2p}/{\partial n^2}$ and
$p_{\varepsilon n}:={\partial^2p}/{\partial\varepsilon\,\partial n}$.

The local curvature perturbation, ${^3\!}R_{\een}$, caused by a local
density perturbation as well as the divergence, ${\vartheta}_{\een}$,
of the local fluid velocity caused by local pressure gradients, are
incorporated into system~(\ref{subeq:final}) via
expressions~(\ref{eq:sol-3R1}) and~(\ref{eq:sol-theta1}),
respectively.

The system of equations~(\ref{subeq:final}) allows us to study the
evolution of the relative density perturbations~(\ref{eq:contrast}) in
closed $K=+1$, flat $K=0$, and open $K=-1$ \textsc{flrw} universes,
where ${^3\!}R_\nul=-6K/a^{2}$.

Sec.\,\ref{sec:flat-flrw-universe} will demonstrate that $p_{n}<0$
before and after the decoupling of radiation and matter.  For
$p_{n}>0$, however, the coefficient in Eq.\,(\ref{fir-ord}) is
positive, causing $\delta_{n}-\delta_{\varepsilon}/(1+w)$ to grow
exponentially, which is not physically plausible.

In the system~(\ref{subeq:final}), Eq.\,(\ref{sec-ord}) is the
evolution equation for energy density perturbations.  Since $p_{n}<0$
for an equation of state $p=p(n,\varepsilon)$, the right-hand side of
Eq.\,(\ref{sec-ord}) is nonzero.  This indicates that density
perturbations are nonadiabatic.  Density perturbations are adiabatic
when the source term of Eq.\,(\ref{sec-ord}) vanishes, i.e., when
\begin{equation}
  \label{eq:adiabatic-condition}
  \delta_n - \frac{\delta_\varepsilon}{1 + w}=0.
\end{equation}
Therefore, Eq.\,(\ref{fir-ord}) can be regarded as the entropy
equation.  An explanation is provided by
Sec.\,\ref{subsec:entropy-perturbation}.

Since $p_{n}<0$, the entropy equation~(\ref{fir-ord}) shows that
\begin{equation}
  \label{eq:dn-enw-klein}
  \left|\delta_{n}-\dfrac{\delta_{\varepsilon}}{1+w}\right|
        \rightarrow 0.
\end{equation}
This means that relative perturbations in energy density are coupled
with those in particle number density.  Since general relativity is
indifferent to the composition of matter, this is true for
\textsc{cdm} as well as for ordinary matter, where it is assumed that
\textsc{cdm} interacts only via gravity.  Therefore, the contraction
of \textsc{cdm} before decoupling was impossible.  This precludes the
formation of potential wells into which ordinary matter could fall
after decoupling to form structures.  Consequently, the initiation of
structure formation by \textsc{cdm} after decoupling is also
impossible.

\section{Thermodynamic Quantities}
\label{sec:thermodyn}

To link an equation of state to Eqs.\,(\ref{subeq:final}), one needs
gauge-invariant expressions for the entropy, temperature, and pressure
perturbations.  These quantities must be expressed in terms of the
relative density perturbations~$\delta_{\varepsilon}$
and~$\delta_{n}$.  This section presents the derivation of these
expressions.

\subsection{Entropy Perturbations}
\label{subsec:entropy-perturbation}

The combined first and second laws of thermodynamics for a simple,
single-species system is given by (see, for example,
Ref.\,\cite{Andersson2021}, Sec.\,2.1)
\begin{equation}
  \label{eq:combined-fs-thermo}
  {\dif}E=T{\dif}S-p\,{\dif}V+\mu\,{\dif}N,
\end{equation}
where $E$, $S$, and $N$ are the energy, the entropy and the number of
particles of a system with volume $V$, temperature $T$, and
pressure~$p$.  The thermal --- or chemical --- potential $\mu$, is the
energy required to add one particle to the system.  In terms of the
particle number density $n=N/V$, the energy per particle $E/N=\varepsilon/n$
and the entropy per particle $s=S/N$ the law
(\ref{eq:combined-fs-thermo}) can be rewritten in the form
\begin{equation}
  \label{eq:law-rewritten}
  {\dif}\left(\dfrac{\varepsilon}{n}N\right)=T{\dif}(sN)-
    p\,{\dif}\left(\dfrac{N}{n}\right)+\mu\,{\dif}N,
\end{equation}
where $\varepsilon$ is the energy density.  The system is
\emph{extensive}, i.e.,
$S(\lambda E, \lambda V, \lambda N)=\lambda S(E,V,N)$, which implies
that the entropy of the gas is given by $S=(E+pV-\mu N)/T$.  Upon
dividing this relation by $N$ the Euler relation is obtained:
\begin{equation}
  \label{eq:chemical-pot}
  \mu=\dfrac{\varepsilon+p}{n}-Ts.
\end{equation}
Eliminating $\mu$ in~(\ref{eq:law-rewritten}) with the aid
of~(\ref{eq:chemical-pot}) reveals that the combined first and second
laws of thermodynamics~(\ref{eq:combined-fs-thermo}) can be expressed
in a form that does not include $\mu$ and $N$ (see
Ref.\,\cite{2001CQGra..18.3917M}):
\begin{equation}
  \label{eq:thermo}
  T{\dif} s= {\dif}\Bigl(\dfrac{\varepsilon}{n}\Bigr)+
      p\,{\dif}\Bigl(\dfrac{1}{n}\Bigr).
\end{equation}
From the background equations~(\ref{subeq:einstein-flrw}) and the
thermodynamic law~(\ref{eq:thermo}) it can be demonstrated that
${\dot{s}_\nul=0}$, indicating that the expansion of the universe
takes place without generating entropy.  From the definition of
gauge-invariant quantities~(\ref{eq:alg-def-gi}), it can be concluded
that $s_\een=s_\een^\phys$ is automatically gauge-invariant.  By
using~(\ref{eq:equal-gi-non-gi}) and $w:=p_\nul/\varepsilon_\nul$, the
thermodynamic relation~(\ref{eq:thermo}) can be rewritten as
\begin{equation}
  \label{eq:thermo-n0-e0-inv}
  T_\nul s^\phys_\een=-\dfrac{\varepsilon_\nul(1+w)}{n^2_\nul}\left(n^\phys_\een-
       \dfrac{n_\nul}{\varepsilon_\nul(1+w)}\varepsilon^\phys_\een\right).
\end{equation}
By using the definitions~(\ref{eq:contrast}), the following is
obtained:
\begin{equation}
  \label{eq:thermo-een}
   T_\nul s^\phys_\een=-\dfrac{\varepsilon_\nul(1+w)}{n_\nul}
   \left(\delta_n-\dfrac{\delta_\varepsilon}{1+w}\right).
\end{equation}
Therefore, Eq.\,(\ref{fir-ord}) can be regarded as the evolution
equation for the entropy per particle.

\subsection{Temperature and Pressure Perturbations}

Since the background temperature $T_{\nul}$ and pressure $p_{\nul}$
are scalars, their gauge-invariant perturbations are defined using the
definition~(\ref{eq:alg-def-gi}):
\begin{equation}
   \label{eq:gi-temp-press}
      T^\phys_\een : =T_\een-\dfrac{\dot{T}_\nul}
      {\dot{\theta}_\nul}\theta_\een,  \quad
     p^\phys_\een  :=p_\een-\dfrac{\dot{p}_\nul}
      {\dot{\theta}_\nul}\theta_\een.
\end{equation}
The expression for $T^\phys_\een$ will be used in
Sec.\,\ref{subsec:mat-dom-era-flat}\@.  In order to arrive at the
gauge-invariant counterpart of~(\ref{eq:p1-a}), it is first necessary
to eliminate $\varepsilon_{\een}$ and $n_{\een}$ from~(\ref{eq:p1-a})
using~(\ref{e-n-gi}). Using
$\dot{p}_{\nul}=p_{n}\dot{n}_{\nul}+
p_{\varepsilon}\dot{\varepsilon}_{\nul}$ and~(\ref{eq:gi-temp-press}),
yields
\begin{equation}
  \label{eq:inv-pressure}
  p^\phys_\een=p_nn^\phys_\een+p_\varepsilon\varepsilon^\phys_\een.
\end{equation}
Eliminating $p_\varepsilon$ with the aid of~(\ref{eq:beta-matter}) and
using the definitions provided in~(\ref{eq:contrast}) for relative
density perturbations results in the following expression:
\begin{equation}
  \label{eq:p-dia-adia}
  p_\een^\phys=\beta^2\varepsilon_\nul\delta_\varepsilon+n_\nul
  p_n\left(\delta_n-\dfrac{\delta_\varepsilon}{1+w}  \right).
\end{equation}
The first term represents the adiabatic component of the pressure
perturbation, while the second term denotes the nonadiabatic
component.

\section{Application: Flat FLRW Universe}
\label{sec:flat-flrw-universe}

The evolution equations presented in~(\ref{subeq:final}), are
applicable to open, flat, and closed \textsc{flrw} universes.  We will
now focus on studying density perturbations in a flat \textsc{flrw}
universe, for which ${^3\!}R_\nul=0$.

Given that the initial density perturbations occurred in the early
universe when $\Lambda\ll\kappa\varepsilon_\nul$, the cosmological
constant $\Lambda$ will be neglected. Consequently, the background
equations~(\ref{subeq:einstein-flrw}), are reduced to
\begin{subequations}
  \label{subeq:einstein-flrw-flat}
\begin{alignat}{3}
  3H^2 & =\kappa\varepsilon_\nul,
         \quad & \kappa& = 8\pi G_{\text{N}}/c^4, & \label{eq:energy-density-constraint-flat}\\
  \dot{\varepsilon}_\nul & = -3H\varepsilon_\nul(1+w), & 
          \quad w & := p_\nul/\varepsilon_\nul, & \label{eq:energy-cons-law-flat} \\
  \dot{n}_\nul & = -3Hn_\nul. & & &  \label{eq:particle-cons-law-flat}
\end{alignat}
\end{subequations}
By using the Friedmann
equation~(\ref{eq:energy-density-constraint-flat}), the
coefficients~(\ref{subeq:coeff-contrast}) of Eq.\,(\ref{sec-ord})
result in
\begin{subequations}
\label{subeq:coeff-contrast-flat}
 \begin{align}
  b_1  & =  H(1-3w-3\beta^2)-2\dfrac{\dot{\beta}}{\beta}, \\
  b_2 & = \kappa\varepsilon_\nul\Bigl(2\beta^2(2+3w)-\tfrac{1}{6}(1+18w+9w^2) \Bigr) \nonumber\\
     &  +2H\dfrac{\dot{\beta}}{\beta}(1+3w)-\beta^2\dfrac{\nabla^2}{a^2}, \\
  b_3 & = \Biggl[\dfrac{-2}{1+w}
  \Biggl(\varepsilon_\nul p_{\varepsilon n}(1+w)
   +\dfrac{2p_n}{3H}\dfrac{\dot{\beta}}{\beta} \nonumber\\
     & +p_n(p_\varepsilon-\beta^2)+n_\nul p_{nn}\Biggr)+
   p_n\Biggr]\dfrac{n_\nul}{\varepsilon_\nul}\dfrac{\nabla^2}{a^2},
\label{eq:b3-flat}
\end{align}
\end{subequations}
where $\nabla^2$ represents the conventional Laplace operator.

\subsection{Era before Decoupling of Matter and Radiation}
\label{subsec:radiation-dom-flat}

In this era the primordial fluid is a mixture of radiation and matter,
wherein the contribution of matter to the pressure is negligible.
Consequently, the equations of state are given by [see
Ref.\,\cite{kodama1984}, {\S\,V-1}]:
\begin{equation}
  \label{eq:state-rad}
  \varepsilon=a_{\text{B}}T_\gamma^4+nmc^{2}, \quad
     p=\tfrac{1}{3}a_{\text{B}}T_\gamma^4.
\end{equation}
The black body constant is represented by $a_{\text{B}}$, the
radiation temperature is denoted by $T_{\gamma}$ and the particle number
density of ordinary matter or \textsc{cdm} is given by~$n$.  Upon
eliminating~$T_\gamma$, the following result is obtained [see
Ref.\,\cite{c8}, Eq.\,(2.10.27)]:
\begin{equation}
  \label{eq:equ-of-state-pressure-rad}
  p=\tfrac{1}{3}(\varepsilon-nmc^{2}).
\end{equation}
By making use of~(\ref{eq:p1-b}), one has
\begin{equation}
  \label{eq:pn-pe-rad}
    p_{n}=-\tfrac{1}{3}mc^{2}, \quad p_{\varepsilon}=\tfrac{1}{3}.
\end{equation}
Since $p_n < 0$, energy density perturbations are coupled to particle
number density perturbations according to (\ref{eq:dn-enw-klein}).

For the parameter $w:=p_{\nul}/\varepsilon_{\nul}$ and the
nonadiabatic speed of sound $\beta$, (\ref{eq:beta-matter}), it is found
\begin{equation}
  \label{eq:beta-2-and-w}
   w=\frac{\tfrac{1}{3}a_{\text{B}}T_{\nul\gamma}^4}
     {a_{\text{B}}T_{\nul\gamma}^4+n_{\nul}mc^{2}}, \quad
  \beta^{2}=\frac{1}{3}-\frac{\tfrac{1}{3}n_{\nul}mc^{2}}
      {\tfrac{4}{3}a_{\text{B}}T_{\nul\gamma}^4+n_{\nul}mc^{2}}.
\end{equation}
The universe was radiation-dominated when
\begin{equation}
  \label{eq:rad-dom-T4-gt-mat}
  a_{\text{B}}T_{\nul\gamma}^4\gg n_{\nul}mc^{2}.
\end{equation}
In this case one has $\beta^2\approx\tfrac{1}{3}$, and
$w\approx\tfrac{1}{3}$, implying that $\dot{\beta}\approx0$.  Upon
substituting these values and expressions~(\ref{eq:pn-pe-rad}) into
the coefficients~(\ref{subeq:coeff-contrast-flat}), and
using~(\ref{eq:rad-dom-T4-gt-mat}) the result is
\begin{equation}
  \ddot{\delta}_\varepsilon-H\dot{\delta}_\varepsilon-
  \left(\dfrac{1}{3}\dfrac{\nabla^2}{a^2}-
    \tfrac{2}{3}\kappa\varepsilon_\nul\right)\delta_\varepsilon=0.
  \label{eq:delta-rad}
\end{equation}
In order to solve Eq.\,(\ref{eq:delta-rad}), it is necessary to use
the solutions to the background
equations~(\ref{subeq:einstein-flrw-flat}). These solutions are given
by
\begin{equation}
  \label{eq:exact-sol-rad}
  H=\tfrac{1}{2}(ct)^{-1}, \quad \kappa\varepsilon_\nul=\tfrac{3}{4} (ct)^{-2},
       \quad a\propto t^{1/2}.
\end{equation}
The dimensionless time, denoted by the symbol~$\tau$, is defined by
$\tau:= t/t_0\ge1$.  This definition implies that
\begin{equation}
  \dfrac{{\dif}^k}{c^k{\dif}t^k}=
  \left(\dfrac{1}{ct_0}\right)^k\dfrac{{\dif}^k}{{\dif}\tau^k}=
   \bigl(2H(t_0)\bigr)^k
   \dfrac{{\dif}^k}{{\dif}\tau^k}.  \label{dtau-n}
\end{equation}
By using the Helmholtz equation
$\nabla^{2}\delta_{\varepsilon}=-|\bm{q}|^{2}\delta_{\varepsilon}$ and
the result derived in~(\ref{dtau-n}), it can be shown that the
equation~(\ref{eq:delta-rad}) can be rewritten as
\begin{equation}
  \label{eq:new-rad}
  \delta_\varepsilon^{\prime\prime}-\dfrac{1}{2\tau}\delta_\varepsilon^\prime+
  \left(\dfrac{\mu_{\text{r}}^2}{4\tau}+
    \dfrac{1}{2\tau^2}\right)\delta_\varepsilon=0,
\end{equation}
where the prime denotes differentiation with respect to
the~$\tau$. With $|\bm{q}|=2\pi/\lambda_0$, the parameter
$\mu_{\text{r}}$ is defined~by
\begin{equation}
     \mu_\text{r} :=
     \dfrac{2\pi}{\lambda_0}\dfrac{1}{H(t_0)}\dfrac{1}{\sqrt{3}}.
\label{xi}
\end{equation}
In this expression, $\lambda_0:=\lambda a(t_0)$ represents the
physical scale of a density perturbation at $t=t_0$.

In order to solve Eq.\,(\ref{eq:new-rad}), it is first necessary to
replace the variable $\tau$ with the variable defined by
$x:=\mu_{\text{r}}\sqrt{\tau}$. Upon transforming back to $\tau$, the
general solution of Eq.\,(\ref{eq:new-rad}) with constants of
integration $A_1(\bm{q})$ and $A_2(\bm{q})$ is obtained:
\begin{equation}
 \delta_\varepsilon(\tau,\bm{q}) =
      \bigl[A_1(\bm{q})\sin\left(\mu_{\text{r}}\sqrt{\tau}\right) +
A_2(\bm{q})\cos\left(\mu_{\text{r}}\sqrt{\tau}\right)\bigr]\sqrt{\tau}.
\label{nu13}
\end{equation}
Since $\vartheta_{\een}$ is included in Eq.\,(\ref{eq:delta-rad}) it
yields oscillating relative density perturbations with an increasing
amplitude. In contrast, the standard
equation~(\ref{eq:standard-rad-dom}), for which $\vartheta_{\een}=0$
(i.e., no fluid flow) yields oscillating solutions with a constant
amplitude.

In the case of large-scale perturbations, that is,
${\lambda_0\rightarrow\infty}$, which leads to
$\mu_{\text{r}}\rightarrow0$, (\ref{xi}), the general solution to
Eq.\,(\ref{eq:new-rad}) is given by
\begin{equation}
    \label{delta-H-rad}
    \delta_\varepsilon(\tau) = -\bigl[\delta_\varepsilon(1)-
    2\delta^\prime_\varepsilon(1)\bigr]\tau
   +2\bigl[\delta_\varepsilon(1)
     -\delta^\prime_\varepsilon(1)\bigr]\tau^{1/2}.
\end{equation}
A comparison of the solutions~(\ref{delta-H-rad})
and~(\ref{eq:cont-eq-sol-phys-gauge-A}) reveals that the solution
proportional to~$\tau$ in~(\ref{delta-H-rad}) is also a solution to
the homogeneous part of equation~(\ref{eq:delta-standard-genrel}). The
solution proportional to~$\tau^{1/2}$ is the particular solution to
the inhomogeneous equation. This solution can only be obtained if the
covariant divergence $\vartheta_{\een}$ is taken into account, as
demonstrated in Appendix\;\ref{subsec:rad-dom-ijk}.  The minus sign
preceding the solution proportional to~$\tau$ is a consequence of the
fact that ${\vartheta_{\een}^{\phys} > 0}$, as follows
from~(\ref{eq:cont-eq-sol-phys-gauge-B}).  The presence of the gauge
function $\psi(\bm{x})$ precludes the derivation of the
solution~(\ref{delta-H-rad}) from~(\ref{subeq:standard}).

\subsection{Era after Decoupling of Matter and Radiation}
\label{subsec:mat-dom-era-flat}

Once protons and electrons have combined to form hydrogen, radiation
pressure will be negligible and the equations of state will be those
of a nonrelativistic, monatomic, perfect gas with three degrees of
freedom.  Therefore, pressure can be neglected in the background
universe.  However, the momentum conservation
law~(\ref{eq:mom2-bw-nul}) indicates that, in order to study the
evolution of density perturbations, one must consider pressure
perturbations, since density perturbations are
static~(\ref{eq:non-rel-lim}) in a pressureless fluid.  According to
the literature [see Ref.\,\cite{2001CQGra..18.3917M}, Eq.\,(13), and
Ref.\,\cite{c8}, Eqs.\,(15.8.20)--(15.8.21)], we use the following
equations of state for a nonrelativistic gas:
\begin{align}
   \label{state-mat}
  \varepsilon(n,T) = nmc^2+\tfrac{3}{2}nk_{\text{B}}T, \quad
  p(n,T) = nk_{\text{B}}T.
\end{align}
In these expressions, the symbol $k_{\text{B}}$ represents Boltzmann's
constant, the quantity $m$ denotes the proton rest mass, and the
temperature of the matter is indicated by the symbol $T$. The rest
mass energy density is $nmc^{2}$ and the kinetic energy density is
$\tfrac{3}{2}nk_{\text{B}}T$.

The elimination of the temperature $T$
from the equations of state~(\ref{state-mat}), yields the equation of
state for the pressure [see Ref.\,\cite{c8}, Eq.\,(2.10.27)]:
\begin{equation}
  \label{eq:pne-eq-of-state}
  p(n,\varepsilon)=\tfrac{2}{3}(\varepsilon-nmc^2).
\end{equation}
The partial derivatives are determined using the
expressions~(\ref{eq:p1-b}) and are given by
\begin{equation}
  \label{eq:pne-dpdedpdn}
   p_n=-\tfrac{2}{3}mc^2,  \quad p_\varepsilon=\tfrac{2}{3}.
\end{equation}
The parameter $w$ is defined by
\begin{equation}
  \label{eq:para-w}
  w:=\dfrac{p_\nul}{\varepsilon_\nul}=
  \dfrac{k_{\text{B}}T_\nul}
  {mc^{2}+\tfrac{3}{2}k_{\text{B}}T_\nul}
  \approx\dfrac{k_{\text{B}}T_\nul}{mc^2} \ll 1.
\end{equation}
Upon substituting~(\ref{state-mat}), (\ref{eq:pne-dpdedpdn})
and~(\ref{eq:para-w}) into~(\ref{eq:beta-matter}), one arrives at the
well-known result [see Ref.\,\cite{c8}, Eq.\,(15.8.22)]:
\begin{equation}
  \label{eq:para-beta}
    \quad
     \beta\approx \dfrac{v_{\text{s}}}{c}=\sqrt{\dfrac{5}{3}
       \dfrac{k_{\text{B}}T_\nul}{mc^2}},
\end{equation}
where $v_{\text{s}}$ represents the adiabatic speed of sound.

From $\beta^2\approx\tfrac{5}{3}w$, we find from
Eq.\,(\ref{eq:time-w}) that $\dot{w}\approx-2Hw$. Therefore, with
$H:=\dot{a}/a$, it can be concluded that $w\propto a^{-2}$.  Given
that~$w$ is proportional to $T_{\nul}$, the well-known result (see
Ref.\,\cite{c8}, Eq.\,(15.5.16) with $\gamma=\tfrac{5}{3}$, where
${\gamma:=c_{\text{p}}/c_{\text{v}}}$ is the adiabatic index) is
obtained:
\begin{equation}
  \label{eq:T-nul-propto-a-2}
  T_\nul\propto a^{-2}.
\end{equation}

The subsequent step is to derive the evolution equations for relative
density perturbations.  The
proportionality~(\ref{eq:T-nul-propto-a-2}) implies
with~(\ref{eq:para-beta}) that $\dot{\beta}/\beta=-H$.  The system of
equations~(\ref{subeq:final}) with
coefficients~(\ref{subeq:coeff-contrast-flat}) can now be rewritten in
the following form:
\begin{subequations}
  \label{final-dust}
  \begin{align}
   & \ddot{\delta}_\varepsilon + 3H\dot{\delta}_\varepsilon-
  \left(\beta^2\dfrac{\nabla^2}{a^2}+
     \tfrac{5}{6}\kappa\varepsilon_\nul\right)\delta_\varepsilon=
     -\dfrac{2}{3}\dfrac{\nabla^2}{a^2}
     \left(\delta_n-\delta_\varepsilon\right), \label{dde-dn-de}\\
   & \dfrac{1}{c}\dfrac{\dif}{{\dif} t}
   \left(\delta_n-\delta_\varepsilon\right)=
   -2H\left(\delta_n-\delta_\varepsilon\right), \label{eq:dn-de}
  \end{align}
\end{subequations}
where it is used that $w$ and $\beta^2$ are negligible with respect to
constants of order one.

Given that the term $(\delta_n-\delta_\varepsilon)$ occurs within the
source term of Eq.\,(\ref{dde-dn-de}), it is necessary to first solve
for the solution to Eq.\,(\ref{eq:dn-de}).  By making the substitution
$H := \dot{a}/a$, the following result is obtained:
\begin{equation}
  \label{eq:dn-dn-a-2}
  (\delta_n-\delta_\varepsilon) \propto a^{-2}.
\end{equation}
In order to relate this proportionality to thermodynamic quantities,
it is necessary to express the perturbed equation of state for the
energy density in terms of relative density perturbations.  It can be
deduced from $\varepsilon=\varepsilon(n,T)$ that
\begin{subequations}
  \label{eq:e1-dot-e0}
\begin{align}
  & \dot{\varepsilon}_\nul=\left(\dfrac{\partial\varepsilon}{\partial n}
  \right)_{T} \dot{n}_\nul+
    \left(\dfrac{\partial\varepsilon}{\partial T} \right)_{n}\dot{T}_\nul,
    \label{eq:de1-phys-T} \\
 & \varepsilon_\een=\left(\dfrac{\partial\varepsilon}{\partial n}
  \right)_{T}  n_\een+
   \left(\dfrac{\partial\varepsilon}{\partial T} \right)_{n} T_\een.
   \label{eq:de1-nonphys}
\end{align}
\end{subequations}
By multiplying~(\ref{eq:de1-phys-T}) by
$\theta_\een/\dot{\theta}_\nul$ and subtracting the result
from~(\ref{eq:de1-nonphys}), one obtains the following result:
\begin{equation}
  \label{eq:eps-phys-T}
  \varepsilon_{\een}^{\phys}=\left(\dfrac{\partial\varepsilon}{\partial n}
  \right)_{T}  n_\een^{\phys}+
   \left(\dfrac{\partial\varepsilon}{\partial T} \right)_{n} T_\een^{\phys},
\end{equation}
where expressions~(\ref{e-n-gi}) and~(\ref{eq:gi-temp-press}) have
been used. By using the expression for $\varepsilon$
in~(\ref{state-mat}) to eliminate the partial derivatives, one obtains
the perturbed equation of state:
\begin{equation}
  \label{eq:gauge-dep-e1}
  \varepsilon^\phys_\een=n^\phys_\een mc^2+\tfrac{3}{2}n^\phys_\een k_{\text{B}}T_\nul+
            \tfrac{3}{2}n_\nul k_{\text{B}}T^\phys_\een.
\end{equation}
Upon dividing~(\ref{eq:gauge-dep-e1}) by
$\varepsilon_{\nul} =
n_{\nul}mc^2+\tfrac{3}{2}n_{\nul}k_{\text{B}}T_{\nul}$ and using the
\emph{exact} value of the ratio $w:=p_{\nul}/\varepsilon_{\nul}$
(\ref{eq:para-w}), the perturbed equation of state for the energy
density expressed in relative density
perturbations~(\ref{eq:contrast}) is obtained:
\begin{equation}
  \label{eq:dn-de-dT}
  \delta_n(t,\bm{x}) - \delta_\varepsilon(t,\bm{x})=
        -\tfrac{3}{2}w(t)\delta_T(t,\bm{x}).
\end{equation}
The quantity $\delta_{T}$ is defined by
$\delta_T := T^\phys_\een/T_\nul$.  The solution to
Eq.\,(\ref{eq:dn-de}) is obtained by combining~(\ref{eq:dn-dn-a-2})
and~(\ref{eq:dn-de-dT}).  Using the \emph{approximate} value of $w$
implies that $w\propto a^{-2}$ [see (\ref{eq:T-nul-propto-a-2})],
making $\delta_{T}$ independent of time.  Therefore, the solution to
Eq.\,(\ref{eq:dn-de}) is given by
\begin{equation}
  \label{eq:dT-constant}
  \delta_n(t,\bm{x})-\delta_\varepsilon(t,\bm{x})\approx
       -\tfrac{3}{2}w(t)\delta_T(t_{\text{dec}},\bm{x}).
\end{equation}
The quantity $\delta_T(t_{\text{dec}},\bm{x})$ represents the relative
temperature perturbation of matter at the time $t_{\text{dec}}$, which
marks the decoupling of matter from radiation.

According to~(\ref{eq:thermo-een}), (\ref{eq:para-w})
and~(\ref{eq:dT-constant}) the entropy per particle is:
\begin{equation}
  \label{eq:heat-exchange}
  s^\phys_\een(t,\bm{x}) \approx
  \tfrac{3}{2}k_{\text{B}}\delta_T(t_{\text{dec}},\bm{x}).
\end{equation}
Thus, the quantity $\delta_{T}$ is \emph{random}.  As a result, it is
generally not equal to zero, which implies that density perturbations
are generally nonadiabatic.

Using~(\ref{eq:pne-dpdedpdn}), (\ref{eq:para-beta}), and
(\ref{eq:dT-constant}), the local relative pressure perturbation
$\delta_p:= p^\phys_\een/p_\nul$ can be calculated from
expression~(\ref{eq:p-dia-adia}). One obtains
\begin{equation}
  \label{eq:rel-press-pert}
  \delta_p(t,\bm{x}) \approx \tfrac{5}{3}\delta_\varepsilon(t,\bm{x})+
  \delta_T(t_{\text{dec}},\bm{x}).
\end{equation}
In this context, the terms $\tfrac{5}{3}\delta_\varepsilon(t,\bm{x})$
and $\delta_T(t_{\text{dec}},\bm{x})$ represent the adiabatic and the
random, nonadiabatic relative pressure perturbation, respectively.

The solution~(\ref{eq:dT-constant}) to Eq.\,(\ref{eq:dn-de}) allows
for the reformulation of the second-order equation~(\ref{dde-dn-de}).
To that end, the solutions to the background
equations~(\ref{subeq:einstein-flrw-flat}) are required. Given that
the pressure with respect to the rest mass energy density can be
disregarded ($w\ll1$), the following solutions are obtained:
\begin{equation}
  \label{eq:exact-sol-mat}
  H=\tfrac{2}{3}(ct)^{-1}, \quad \kappa\varepsilon_\nul= \tfrac{4}{3}(ct)^{-2},
      \quad a\propto t^{2/3}.
\end{equation}
The dimensionless time, denoted by the symbol~$\tau$, is defined by
$\tau:= t/t_{\text{dec}}\ge1$.  This definition implies that
\begin{equation}
  \dfrac{{\dif}^k}{c^k{\dif}t^k}=
    \left(\dfrac{1}{ct_{\text{dec}}}\right)^k\dfrac{{\dif}^k}{{\dif}\tau^k}=
   \bigl(\tfrac{3}{2}H(t_{\text{dec}})\bigr)^k
   \dfrac{{\dif}^k}{{\dif}\tau^k}.    \label{dtau-n-dust}
\end{equation}
By using Eqs.\,(\ref{eq:energy-density-constraint-flat})
and~(\ref{eq:dT-constant}), as well as the
expressions~(\ref{eq:para-beta}), (\ref{eq:exact-sol-mat}),
(\ref{dtau-n-dust}), the Helmholtz equation
$\nabla^{2}\delta=-|\bm{q}|^{2}\delta$, and $w\propto a^{-2}$, it can
be demonstrated that Eq.\,(\ref{dde-dn-de}) can be expressed in the
following manner:
\begin{equation}
  \label{eq:eerste-dust-dimless}
   \delta_\varepsilon^{\prime\prime}+\dfrac{2}{\tau}\delta_\varepsilon^\prime+
\left(\dfrac{4}{9}\dfrac{\mu_{\text{m}}^2}{\tau^{8/3}}-\dfrac{10}{9\tau^2}
\right)\delta_\varepsilon=-\dfrac{4}{15}\dfrac{\mu^2_{\text{m}}}{\tau^{8/3}}\delta_T(t_{\text{dec}},\bm{q}),
\end{equation}
where the prime denotes differentiation with respect to~$\tau$. With
$\lvert\bm{q}\rvert=2\pi/\lambda_{\text{dec}}$, the parameter
$\mu_{\text{m}}$ is defined by
\begin{equation}\label{eq:const-mu}
  \mu_{\text{m}}:=\dfrac{2\pi}{\lambda_{\text{dec}}}\dfrac{1}{H(t_{\text{dec}})}
  \dfrac{v_{\text{s}}(t_{\text{dec}})}{c}.
\end{equation}
In this expression, $\lambda_{\text{dec}}:=\lambda a(t_{\text{dec}})$
represents the physical scale of a perturbation at $t=t_{\text{dec}}$,
and $v_{\text{s}}(t_{\text{dec}})/c\approx\beta(t_{\text{dec}})$ is
given by~(\ref{eq:para-beta}).

Finally, the expression~(\ref{eq:rel-press-pert}) enables the
reformulation of Eq.\,(\ref{eq:eerste-dust-dimless}) in a form
suitable for the study of the evolution of relative density
perturbations:
\begin{equation}
  \label{eq:dust-dimless}
  \delta^{\prime\prime}_{\varepsilon}+
  \frac{2}{\tau}\delta^{\prime}_{\varepsilon}+
  \dfrac{4}{15}\dfrac{\mu_{\text{m}}^2}{\tau^{8/3}}\delta_{p}-
  \dfrac{10}{9\tau^2}\delta_{\varepsilon}=0, \quad
  \tau:= \dfrac{t}{t_{\text{dec}}}\ge1.
\end{equation}
The second term in this equation represents the expansion, the third
term is the pressure term, where $\delta_{p}$ is given
by~(\ref{eq:rel-press-pert}), and the fourth term represents
gravitation.

For large-scale perturbations, i.e., when
$\lambda_{\text{dec}}\rightarrow\infty$, implying that
$\mu_{\text{m}}\rightarrow0$, relative pressure
perturbations~$\delta_p$ played a negligible role in their evolution.
Therefore, the solution to Eq.\,(\ref{eq:dust-dimless}) is
\begin{equation}
     \label{eq:new-dust-53-adiabatic-e}
     \delta_\varepsilon(\tau)  = 
     \bigl[\tfrac{5}{7}\delta_\varepsilon(1)
     +\tfrac{3}{7}\delta^\prime_\varepsilon(1)\bigr]\tau^{2/3}+
     \bigl[\tfrac{2}{7}\delta_\varepsilon(1)-
     \tfrac{3}{7}\delta^\prime_\varepsilon(1)\bigr]\tau^{-5/3}.
\end{equation}
The solution proportional to~$\tau^{-5/3}$ dies out quickly. Thus,
large-scale relative density perturbations evolve nearly
proportionally to~$\tau^{2/3}$, which is close to the
solution~(\ref{eq:fluct-p-is-nul}) for a pressureless fluid.  However,
since $\vartheta_\een^\phys(\tau) =-\tfrac{7}{9}\tau^{-4/3}$,
(\ref{eq:sol-theta-phys}), there is a pressure-gradient-driven fluid
flow.  Therefore, the solution~(\ref{eq:new-dust-53-adiabatic-e}) is
invalid for a pressureless fluid.  As shown in
Sec.\,\ref{sec:newt-limit}, the equations for scalar
perturbations~(\ref{subeq:pertub-gi-flat}) are, for a pressureless
fluid, reduced to a single equation, namely the time-independent
Poisson equation~(\ref{eq:poisson}).

A comparison of the solutions~(\ref{eq:new-dust-53-adiabatic-e})
and~(\ref{eq:sol-74c1c2}) reveals that the solution proportional
to~$\tau^{2/3}$ in~(\ref{eq:new-dust-53-adiabatic-e}) is also a
solution to the homogeneous part of Eq.\,(\ref{eq:alg-mat-dom-delta}).
The solution proportional to~$\tau^{-5/3}$ is the particular solution
to the inhomogeneous equation.  This solution can only be obtained if
the covariant divergence~$\vartheta_{\een}$ is taken into account, as
demonstrated in Appendix\;\ref{app:era-after-decoupling}.  Since in
the standard equation~(\ref{eq:standaard-after-dec})
$\vartheta_{\een}=0$ (i.e., no fluid flow), it does not produce
growing density perturbations.

The presence of the gauge constant~$C$,
precludes the derivation of
solution~(\ref{eq:new-dust-53-adiabatic-e}) from the
system~(\ref{subeq:algemene-vergelijkingen-mat-dom}).

\subsubsection{Cosmological Quantities}
\label{sec:p-obs-q}

\begin{table}[t]
  \caption{Planck satellite results\phantom{y}\label{tab:planck-sat-res}}
   \begin{minipage}{55.0mm}
 \begin{ruledtabular}
 \begin{tabular}{rcl}
    $z(t_{\text{dec}})$ &=& 1090 \\
    $z(t_{\text{eq}})$ &=&   3387 \\
    $cH(t_{\text{p}})$ &=&  $67.66\,\text{km}\,\text{s}^{-1}\text{Mpc}^{-1}$ \\
    $T_{\nul\gamma}(t_{\text{p}})$ &=&  $2.725\,\text{K}$ \\
    $t_{\text{p}}$ &=&   $13.79\,\text{Gyr}$ \\
      $\lvert\delta_{T_\gamma}(t_{\text{dec}},\bm{x})\rvert$  &$\lesssim$ &   $10^{-5}$
 \end{tabular}
\end{ruledtabular}
 \end{minipage}
\end{table}

To investigate the evolution of relative density perturbations using
Eq.\,(\ref{eq:dust-dimless}), the parameter
$\mu_{\text{m}}$~(\ref{eq:const-mu}) must first be expressed in terms
of observable quantities, namely the redshift~$z(t_{\text{dec}})$ at
the time of decoupling, the present value of the Hubble
parameter~$cH(t_{\text{p}})$, and the present value of the temperature
$T_{\nul}(t_{\text{p}})$ of the cosmological fluid.

The redshift $z(t)$ is
defined by the expression
\begin{equation}
  \label{eq:redshift}
  z(t):=\dfrac{a(t_{\text{p}})}{a(t)}-1,
\end{equation}
where $a(t_{\text{p}})$ is the present value of the scale factor.  The
Friedmann equation~(\ref{FRW3}) shows that, for a flat \textsc{flrw}
universe, one may set $a(t_{\text{p}})=1$.  Using the background
solutions~(\ref{eq:exact-sol-mat}), it is possible to express the
relevant parameters of the universe in the redshift:
\begin{subequations}
  \label{eq:handig}
  \begin{align}
  H(t)&=H(t_{\text{p}})\bigl(z(t)+1\bigr)^{3/2}, \label{handig-H} \\
  t&=t_{\text{p}}\bigl(z(t)+1\bigr)^{-3/2}, \label{handig-tijd} \\
    T_{\nul\gamma}(t)&=T_{\nul\gamma}(t_{\text{p}})\bigl(z(t)+1\bigr),
                       \label{handig-temp} \\
    n_{\nul}(t)&=n_{\nul}(t_{\text{p}})\bigl(z(t)+1\bigr)^{3}. \label{handig-n0}
\end{align}
\end{subequations}
In deriving relation~(\ref{handig-temp}), it is used that, subsequent
to decoupling, $T_{\nul\gamma}\propto a^{-1}$.  Upon
substituting~(\ref{eq:para-beta}) into expression~(\ref{eq:const-mu})
and using the relations~(\ref{eq:handig}), we obtain
\begin{equation}
  \label{eq:H-dec-wmap}
  \mu_{\text{m}}=\dfrac{2\pi}{\lambda_{\text{dec}}}
     \dfrac{1}{cH(t_{\text{p}})}\dfrac{1}{z(t_{\text{dec}})+1}
\sqrt{\dfrac{5}{3}\dfrac{k_{\text{B}}T_{\nul\gamma}(t_{\text{p}})}{m}},
\end{equation}
where it is assumed that at decoupling, the matter and radiation
temperatures were equal, that is,
$T_\nul(t_{\text{dec}})=T_{\nul\gamma}(t_{\text{dec}})$.  In this
manner, the parameter~$\mu_{\text{m}}$ has been expressed in
observable quantities.  Upon substituting the numerical values from
Tables~\ref{tab:planck-sat-res} and~\ref{tab:phys-const}, one obtains
\begin{equation}\label{eq:nu-m-lambda}
    \mu_{\text{m}}=\dfrac{16.48}{\lambda_{\text{dec}}}, \quad
\lambda_{\text{dec}} \text{ in pc}.
\end{equation}
It can thus be concluded that the parameter~$\mu_{\text{m}}$ is
dependent solely on the initial scale~$\lambda_{\text{dec}}$ of a
density perturbation.

Finally, the dimensionless time, defined as $\tau:=t/t_{\text{dec}}$,
is expressed in terms of the redshift, given~by:
\begin{equation}
  \label{eq:tau-zdec-zt}
  \tau = \left(\dfrac{z(t_{\text{dec}})+1}{z(t)+1}  \right)^{3/2},
\end{equation}
where relation~(\ref{handig-tijd}) with $z(t_{\text{p}})=0$ is used.

\begin{table}[t]
  \caption{Physical constants\label{tab:phys-const}}
   \begin{minipage}{55.0mm}
 \begin{ruledtabular}
 \begin{tabular}{rcl}
      $m$ &=&  $1.6726\times10^{-27}\,\text{kg}$ \\
                  $ \text{pc} $ &=& $3.0857\times10^{16}\,\text{m}=3.2616\,\text{ly} $ \\
         $c$ &=& $2.9979\times10^8\,\text{m}\,\text{s}^{-1}$ \\
        $  k_{\text{B}}$ &=& $ 1.3806\times10^{-23}\,\text{J}\,\text{K}^{-1} $\\
                  $  G_{\text{N}} $  &=& $6.6743\times10^{-11}\,\text{m}^3\,\text{kg}^{-1}\,\text{s}^{-2}$ \\
         $ \text{M}_\odot$ &=& $1.9889\times10^{30}\,\text{kg}$ \\
         $  a_{\text{B}}$ &=& $7.5657\times10^{-16}\,\text{J}\,\text{m}^{-3}\,\text{K}^{-4}$
 \end{tabular}
\end{ruledtabular}
 \end{minipage}
\end{table}

\subsubsection{Initial Values}
\label{subsubsec:init-vals}

In order to solve Eq.\,(\ref{eq:dust-dimless}), it is necessary to
determine the initial values of the quantities $\delta_\varepsilon$
and $\delta^\prime_\varepsilon$. In addition, values of the local
random, nonadiabatic relative pressure perturbation~$\delta_T$ in
expression~(\ref{eq:rel-press-pert}), are required.

\paragraph{Planck Satellite}

The Planck observations~\cite{2020A&A...641A...6P-verkort} of the
relative perturbations $\delta_{T_\gamma}(t_{\text{dec}},\bm{x})$ in the
background radiation temperature imply that
$\lvert\delta_\varepsilon(t_{\text{dec}},\bm{x})\rvert\lesssim10^{-5}$.
In the absence of knowledge regarding the initial growth rate, it is
assumed that
$\dot{\delta}_\varepsilon(t_{\text{dec}},\bm{x}) \approx 0$.
Therefore, the initial values for Eq.\,(\ref{eq:dust-dimless}) are as
follows:
\begin{equation}
  \label{eq:init-delta-tau}
  \lvert\delta_\varepsilon(t_{\text{dec}},\bm{q})\rvert\lesssim 10^{-5},
     \quad \delta^\prime_\varepsilon(t_{\text{dec}},\bm{q})=0.
\end{equation}
As demonstrated in the calculations presented in
Sec.\,\ref{sec:hier-struc}, the outcome of
Eq.\,(\ref{eq:dust-dimless}) is almost entirely independent of the
initial value for $\delta_\varepsilon$, provided that it satisfies the
condition
$\lvert\delta_\varepsilon(t_{\text{dec}},\bm{q})\rvert\le10^{-4}$.

\paragraph{Random Nonadiabatic Pressure Perturbations}
\label{par:initial-non-adiabtic-pressure}

The initial values $\delta_\varepsilon(t_{\text{dec}},\bm{q})$ and
$\delta_n(t_{\text{dec}},\bm{q})$ are related to the local random,
nonadiabatic relative pressure perturbation
$\delta_{T}(t_{\text{dec}},\bm{q})$ by
expression~(\ref{eq:dT-constant}).  Since ${w\ll1}$, very small values
of the difference
$\delta_{n}(t_{\text{dec}},\bm{q})-\delta_{\varepsilon}(t_{\text{dec}},\bm{q})$
result in relatively large positive or negative values of~$\delta_{T}(t_{\text{dec}},\bm{q})$.

The fact that $\delta_{T}(t_{\text{dec}},\bm{q})$ takes on random
values follows from expression~(\ref{eq:heat-exchange}).  However,
using Eq.\,(\ref{eq:dT-constant}) the randomness of
$\delta_{T}(t_{\text{dec}},\bm{q})$ can also be explained as follows.
The transition of the universe from the era preceding decoupling to
the era following decoupling was rapid and chaotic.  A notable decline
was observed in the mean particle velocity, and the pressure.  As a
consequence, the values of both
$\delta_{\varepsilon}(t_{\text{dec}},\bm{q})$ and
$\delta_{n}(t_{\text{dec}},\bm{q})$ exhibited minor irregularities on
the surface of last scattering.  Therefore, it is probable that after
the transition, the initial values of the difference between the
relative density perturbations, represented by
${(\delta_n-\delta_\varepsilon)(t_{\text{dec}},\bm{q})}$, were
randomly distributed among all density perturbations.

\subsubsection{Structure Formation}
\label{sec:hier-struc}

The subsequent analysis will use Eq.\,(\ref{eq:dust-dimless}) to
examine the evolution of relative density perturbations.  This
equation shows that the evolution of a relative perturbation,
$\delta_{\varepsilon}(t,\bm{q})$, with an initial
scale~$\lambda_{\text{dec}}$, was influenced by three factors: the
expansion of the universe, pressure perturbations, and gravitation.
Relative pressure perturbations~$\delta_{p}$, as indicated
by~(\ref{eq:rel-press-pert}), consisted of two discrete components: an
adiabatic contribution, represented by
$\tfrac{5}{3}\delta_\varepsilon(t,\bm{q})$, and a
random~(\ref{eq:heat-exchange}), nonadiabatic contribution,
represented by $\delta_{T}(t_{\text{dec}},\bm{q})$, where the
adiabatic component was initially
negligible~(\ref{eq:init-delta-tau}).

\begin{figure}
\begin{center}
\includegraphics[width=\columnwidth]{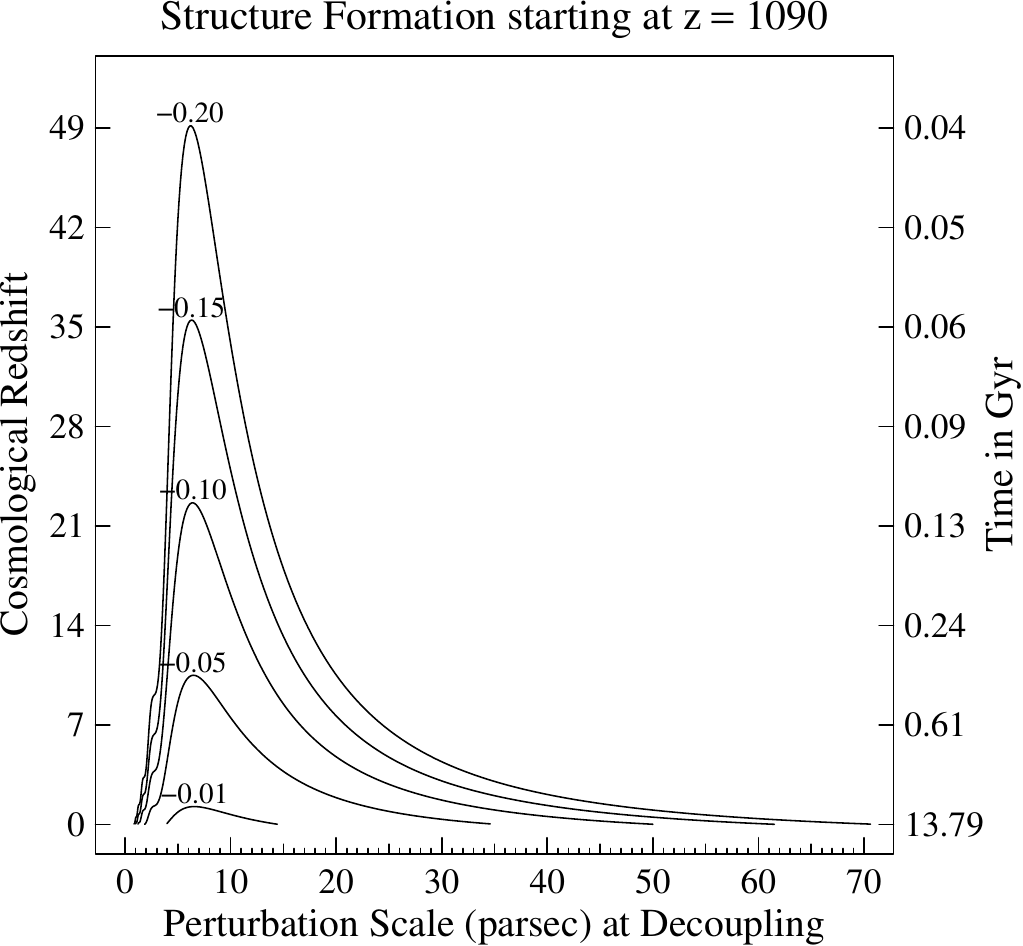}
\caption{The graphs show the redshift and time when a relative
  perturbation in the energy density with initial scale
  $\lambda_{\text{dec}}$, and initial values
  $\delta_\varepsilon(t_{\text{dec}},\bm{q})\approx10^{-5}$ and
  $\delta^\prime_\varepsilon(t_{\text{dec}},\bm{q})=0$ starting to
  grow at an initial redshift of $z(t_{\text{dec}})=1090$ has become
  nonlinear, i.e., when $\delta_\varepsilon(t,\bm{q})=1$.  The graphs
  are labeled with the initial values of the random, nonadiabatic
  pressure perturbations $\delta_T(t_{\text{dec}},\bm{q})$. For each
  graph, the Jeans scale is~$6.4\,\text{pc}$.}
\label{fig:collapse}
\end{center}
\end{figure}

Since $w\ll1$, (\ref{eq:para-w}), the nonadiabatic component could
have taken on relatively large positive or negative values for small
differences in the relative perturbations of energy and particle
number density, as shown in~(\ref{eq:dT-constant}).  Consequently,
there were regions where the total relative pressure perturbation was
initially negative after decoupling. This allowed density
perturbations to grow rapidly, despite the fact that the expansion of
the universe opposed growth. As the adiabatic component of the
relative pressure perturbation increased, the total relative pressure
perturbation became positive. This significantly reduced the growth
rate of density perturbations. The initial growth phase was brief yet
long enough for a density perturbation to reach the nonlinear regime
several hundred million years after the Big Bang. Conversely, if the
nonadiabatic relative pressure perturbation was positive initially,
then the total relative pressure perturbation was also positive. This
resulted in the formation of voids, which were regions with negative
values of~$\delta_{\varepsilon}$. In these regions, matter was driven
to the edges.

Fig.\,\ref{fig:collapse} provides a summary of the evolution of
relative density perturbations. It was created using the following
methodology.  For each value of ${\delta_T=-0.01}$, $-0.05$, $-0.10$,
$-0.15$, and $-0.20$, Eq.\,(\ref{eq:dust-dimless}) was solved
numerically for a large number of values for the initial perturbation
scale $\lambda_{\text{dec}}$ using the initial
values~(\ref{eq:init-delta-tau}).  The integration started at $z=1090$
and ended when either $z=0$ or when $\delta_\varepsilon=1$.  Each
integration run returned a single point on the graph for a particular
choice of the initial scale $\lambda_{\text{dec}}$ when $z>0$ and
$\delta_\varepsilon=1$. In this case, it is clear that the
perturbation has become nonlinear within $13.79\,\text{Gyr}$.  On the
other hand, if the integration stopped at $z=0$ and
$\delta_\varepsilon < 1$, then the perturbation had not yet reached
its nonlinear phase today.  Each graph shows the time and scale at
which $\delta_\varepsilon=1$ for a given value of~$\delta_T$.

\begin{figure}
\begin{center}
\includegraphics[width=\columnwidth]{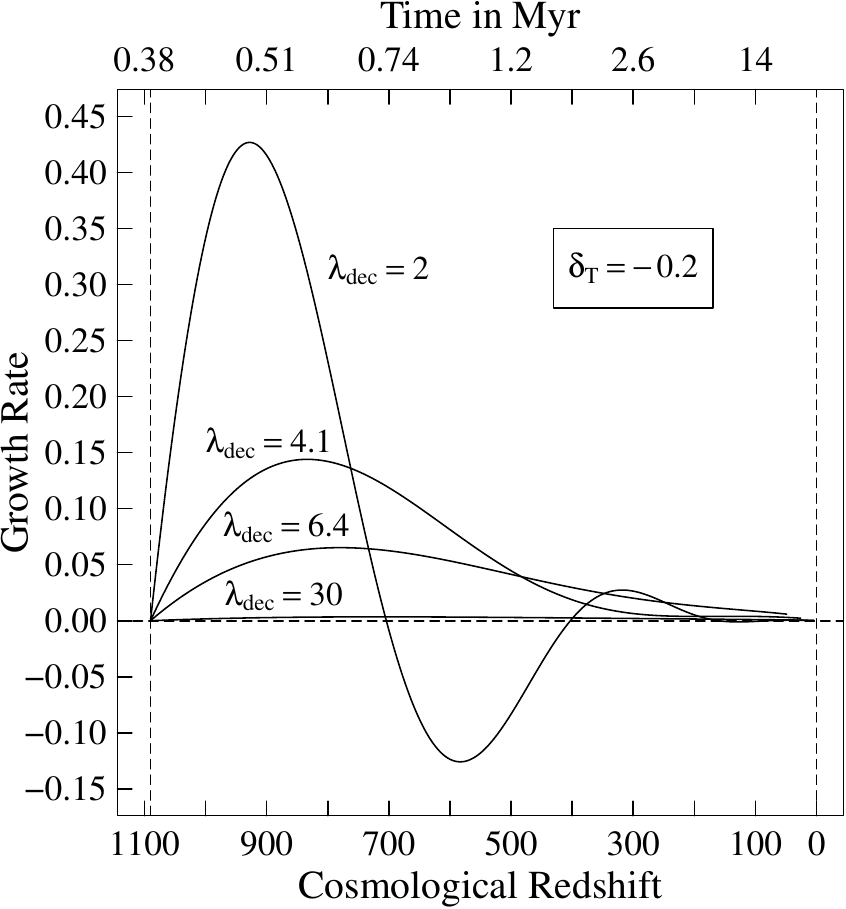}
\caption{The graphs show the growth rates $\delta^\prime_\varepsilon$,
  with initial values
  $\delta_\varepsilon(t_{\text{dec}},\bm{q})\approx10^{-5}$ and
  $\delta^\prime_\varepsilon(t_{\text{dec}},\bm{q})=0$, as function of
  the redshift $z$, or time in million of years.  The initial scales
  $\lambda_{\text{dec}}$ of the perturbations are measured in
  parsec. The evolution of relative density perturbations started at
  $z=1090$.}
\label{fig:growth-rates}
\end{center}
\end{figure}

As illustrated in Fig.\,\ref{fig:collapse}, the optimal scale for
growth was approximately $6.4\,\text{pc}$.  Perturbations at this
scale entered the nonlinear phase as early as $40$ million years
($z\approx49$) after the Big Bang.  Perturbations with scales smaller than
$6.4\,\text{pc}$ reached their nonlinear phase at a much later time
because their internal gravitation was weaker than for large-scale
perturbations.  Furthermore, pressure perturbations and expansion of
the universe led to oscillatory behavior, as shown in
Fig.\,\ref{fig:growth-rates}.  In contrast, perturbations with a scale
greater than $6.4\,\text{pc}$ were less affected by pressure
perturbations. However, because of their larger size, the expansion
worked against their growth, so that, despite their stronger
gravitation, they also reached the nonlinear phase at a later time.
Perturbations larger than $70\,\text{pc}$ did not reach their
nonlinear regime within $13.79\,\text{Gyr}$.

Perturbations that became nonlinear within $13.79\,\text{Gyr}$ are
within the particle horizon at decoupling given~by:
\begin{equation}
  \label{eq:horizon-size}
  d_{\text{H}}(t_{\text{dec}})=
  a(t_{\text{dec}})\int_{0}^{ct_{\text{dec}}}\!\!\dfrac{\dif\tau}{a(\tau)}=
  3ct_{\text{dec}}
      \approx3.5\times10^5\,\text{pc},
\end{equation}
where Eq.\,(\ref{handig-tijd}) and Tables~\ref{tab:planck-sat-res}
and~\ref{tab:phys-const} have been used.

As Fig.\,\ref{fig:growth-rates} shows, the largest initial growth
rates occurred for perturbations that were smaller than
$6.4\,\text{pc}$.  The smallest perturbations which had become
nonlinear within $13.79\,\text{Gyr}$, had a scale of approximately
$2\,\text{pc}$, as shown in Fig.\,\ref{fig:collapse}.  Perturbations
with scales
${2\,{\text{pc}}\lesssim\lambda_{\text{dec}}\lesssim4.1\,\text{pc}}$
oscillated toward the nonlinear phase within $13.79\,\text{Gyr}$.
After approximately $14$~million years, the total pressure
perturbation~(\ref{eq:rel-press-pert}) had become positive and the
growth rate had decreased, resulting in gravitation and, to a lesser
extent, expansion becoming the primary drivers of the evolution of
density perturbations. Consequently, the most turbulent phase of
density perturbations had concluded, and a phase of gradual and
consistent gravitational growth toward the nonlinear regime had
commenced.  As can be seen in Fig.\,\ref{fig:collapse}, at a redshift
of $z\approx7$, or $600$~million years after the Big Bang, all
perturbations with scales between $2\,\text{pc}$ and $24\,\text{pc}$
had become nonlinear.

Assuming a spherical density perturbation with a diameter equal
to~$\lambda_{\text{dec}}$, the mass of the perturbation at decoupling
is given by
\begin{equation}
  \label{eq:M-dec-rel}
  M(t_{\text{dec}})= \dfrac{4\pi}{3}\bigl(\tfrac{1}{2}\lambda_{\text{dec}}
  \bigr)^3  n_\nul(t_{\text{dec}})m.
\end{equation}
The particle number density at the time of decoupling,
$n_\nul(t_{\text{dec}})$, can be derived from the Friedmann
equation~(\ref{eq:energy-density-constraint-flat}).  For $w\ll1$ and
at time $t_{\text{p}}$ this equation is given~by:
\begin{equation}
  \label{eq:friedman-tp}
  3H^{2}(t_{\text{p}})=\kappa n_{\nul}(t_{\text{p}})mc^{2},
  \quad  \kappa = 8\pi G_{\text{N}}/c^4.
\end{equation}
By using the relation~(\ref{handig-n0}), one obtains
\begin{equation}
  \label{eq:M-dec-n-dec}
  n_\nul(t_{\text{dec}})=\frac{3\bigl[cH(t_{\text{p}})\bigr]^{2}
    \bigl[z(t_{\text{dec}})+1\bigr]^{3}}{8\pi m G_{\text{N}}}.
\end{equation}
The relations~(\ref{eq:M-dec-rel}) and~(\ref{eq:M-dec-n-dec}) and the
constants in Tables~\ref{tab:planck-sat-res} and~\ref{tab:phys-const}
reveal that perturbations with scales between $2\,\text{pc}$ and
$24\,\text{pc}$, had masses between $6.7\times10^{2}\,\text{M}_\odot$
and $1.2\times10^6\,\text{M}_\odot$.

Because of the steepness of the graphs in Fig.\,\ref{fig:collapse} for
scales below $6.4\,\text{pc}$, this scale is designated as the
relativistic Jeans scale.  Its value is
\begin{equation}
  \lambda_{\text{Jeans}}(t_{\text{dec}}):=\lambda_{\text{Jeans}}
a(t_{\text{dec}})\approx6.4\,\text{pc}.
\end{equation}
Accordingly, the Jeans mass at decoupling is given by
\begin{equation}
  \label{eq:J-Mass-decoup}
  M_{\text{Jeans}}(t_{\text{dec}})\approx 2.2\times10^4\, \text{M}_\odot.
\end{equation}
The above analysis shows that, if nonadiabatic pressure perturbations
are taken into account, relative density perturbations with initial
values
$\lvert\delta_\varepsilon(t_{\text{dec}},\bm{x})\rvert\lesssim10^{-5}$
and ${\dot{\delta}_\varepsilon(t_{\text{dec}},\bm{x}) \approx 0}$ can
give rise to structure in the early universe after the decoupling of
matter and radiation.

\section{Summary and Conclusion}
\label{sec:conclusion}

This article introduces a relativistic cosmological perturbation
theory~(\ref{subeq:final}) that describes the evolution of relative
energy density and particle number density perturbations in closed
($K=+1$), flat ($K=0$), and open (${K=-1}$)
Friedmann-Lemaître-Robertson-Walker (\textsc{flrw}) universes, where
${^3\!}R_\nul=-6K/a^{2}$ is the spatial curvature.  It consists of two
differential equations: a second-order inhomogeneous evolution
equation for the relative energy density perturbation and a
first-order entropy equation that determines the evolution of the
source term of the first equation.  These equations together describe
the evolution of nonadiabatic relative perturbations
$\delta_{\varepsilon}(t,\bm{x}):=\varepsilon^{\phys}_{\een}(t,\bm{x})/\varepsilon_{\nul}(t)$
and $\delta_{n}(t,\bm{x}):=n^{\phys}_{\een}(t,\bm{x})/n_{\nul}(t)$ in
energy and particle number densities, respectively.

The entropy equation implies that before matter and radiation
decoupled, both ordinary matter perturbations and Cold Dark Matter
(\textsc{cdm}) perturbations are coupled to energy density
perturbations by gravitation.  Consequently, \textsc{cdm} did not
initiate structure formation after decoupling.

After decoupling, pressure gradients became essential for fluid flow, which is crucial for forming structures, since in a pressureless fluid, density perturbations~$\varepsilon^{\phys}_{\een}$ are static.

Newtonian gravity is not a suitable framework for studying the
evolution of cosmological density perturbations.  This is because the
conventional evolution equation, whether derived from general
relativity or Newtonian gravity, yields gauge-dependent solutions.
Consequently, no distinction exists between sub- and super-horizon
density perturbations.

\subsection{Results for a Flat FLRW Universe}

In a radiation-dominated universe, relative density
perturbations~$\delta_{\varepsilon}$ oscillated with an amplitude that
increased proportionally to the square root of time.

After matter and radiation decoupled at $t=t_{\text{dec}}$, the
cosmological fluid became a nonrelativistic perfect fluid.  According
to observations made by the Planck
satellite~\cite{2020A&A...641A...6P-verkort}, relative density
perturbations were of the order of~$10^{-5}$ at the time of
decoupling.  These perturbations are far too small to explain
structures in the universe by gravitational growth, if only adiabatic
perturbations are considered. This is because the growth of an
adiabatic density perturbation slows down due to the increasing
pressure perturbation caused by the density perturbation itself.  For
this reason, this article examines nonadiabatic density perturbations.
This was achieved using an equation of state for a nonrelativistic
perfect gas that considers not only the particles’ rest energy, but
also their kinetic energy.  Therefore, in addition to the energy
density, the particle number density must also be considered. Then,
the total pressure is given by
$\delta_p(t,\bm{x}) \approx \tfrac{5}{3}\delta_\varepsilon(t,\bm{x})+
\delta_T(t_{\text{dec}},\bm{x})$, where
$\tfrac{5}{3}\delta_{\varepsilon}(t,\bm{x})$ is the adiabatic
component caused by the density perturbation itself, and a random
nonadiabatic component~$\delta_{T}(t_{\text{dec}})$.  In addition, the
solution to the entropy equation is identical to the perturbed
equation of state of the energy density, expressed in terms of
relative density perturbations and the nonadiabatic component of
pressure, that is,
$\delta_n(t,\bm{x})-\delta_\varepsilon(t,\bm{x})\approx
-\tfrac{3}{2}w(t)\delta_T(t_{\text{dec}},\bm{x})$, where
$w:=p_{\nul}/\varepsilon_{\nul}\ll1$.  This expression shows that
small random fluctuations in both~$\delta_{n}(t_{\text{dec}},\bm{x})$
and~$\delta_{\varepsilon}(t_{\text{dec}},\bm{x})$ --- arising from the
rapid and chaotic transition to the post-decoupling era --- could lead
to significant local random fluctuations in the relative nonadiabatic
pressure perturbation, $\delta_{T}(t_{\text{dec}},\bm{x})$, which
could take on both positive and negative values.  If the total
relative pressure perturbation~$\delta_{p}(t_{\text{dec}},\bm{x})$ was
positive at the moment of decoupling, a void was formed and matter
moved toward its edges.  Conversely,
if~$\delta_{p}(t_{\text{dec}},\bm{x})$ was negative, the density
perturbation would undergo a brief period of rapid growth.  This rapid
growth continued until the total relative pressure
perturbation~$\delta_{p}(t,\bm{x})$ became positive. After that, the
density perturbation slowly but steadily grew toward the nonlinear
phase, which was reached early on.  Perturbations smaller than
$70\,\text{pc}$ would reach their nonlinear phase within
$13.79$~billion years or less.  The relativistic Jeans scale was
$6.4\,\text{pc}$, and the corresponding Jeans mass was
$2.2\times10^4\,\text{M}_\odot$.  Perturbations of Jeans scale size
reached their nonlinear regime at a redshift of $z\approx49$ or
approximately $40$~million years after the Big Bang.  All density
perturbations with scales between $2\,\text{pc}$ and $24\,\text{pc}$
reached their nonlinear phase within $z\approx7$ or $600$~million
years. These scales correspond to masses between
$6.7\times10^{2}\,\text{M}_\odot$ and $1.2\times10^6\,\text{M}_\odot$.
This is smaller than what was observed by the James Webb Space
Telescope (\textsc{jwst})~\cite{2023arXiv231214738C}, which observed
mature galaxies.  Nevertheless, it is reasonable to assume that the
masses continued to increase in the nonlinear phase due to their
greater gravitational pull.

\subsection{Conclusion}

Investigating cosmological density perturbations in a flat
\textsc{flrw} universe reveals that if a nonrelativistic, monatomic,
perfect gas with three degrees of freedom accurately represents the
cosmological fluid in our universe after matter and radiation
decoupled, then a perturbation theory based on general relativity can
explain the origin of the first structures in the universe without
relying on \textsc{cdm}.  Subsequent observations with the
\textsc{jwst}, launched on Christmas Day $2021$, are expected to yield
novel insights into the conditions that ensued immediately after the
decoupling of radiation and matter, as well as the formation of the
earliest structures in the universe.  These observations will help us
determine if general relativity accurately describes the formation of
the first structures in the universe.




\begin{appendix}

\section{Derivation of the \protect\\ Perturbation Theory}
\label{app:computer-algebra}

\begin{table*}
\caption{The coefficients $\alpha_{ij}$ figuring in
  the equations~(\ref{subeq:nieuw})  \label{eq:aij}}
\begin{minipage}{150.0mm}
\begin{ruledtabular}  
   \begin{tabular}{cccc}
  $3H(1+p_\varepsilon)+\dfrac{\kappa\varepsilon_\nul(1+w)}{2H}$ &
      $3Hp_n$ & $\varepsilon_\nul(1+w)$ &
      $\dfrac{\varepsilon_\nul(1+w)}{4H}$ \\  
 & & & \\
  $\dfrac{\kappa n_\nul}{2H}$ & $3H$ & $n_\nul$ & $\dfrac{n_\nul}{4H}$ \\  
 & & & \\
  $\dfrac{p_\varepsilon}{\varepsilon_\nul(1+w)}\dfrac{\tilde{\nabla}^2}{a^2}$ &
       $\dfrac{p_n}{\varepsilon_\nul(1+w)}\dfrac{\tilde{\nabla}^2}{a^2}$ &
       $H(2-3\beta^2)$ & $0$ \\
 & & & \\
  $\dfrac{\kappa\,{^3\!}R_\nul}{3H}$ & $0$ &
  $-2\kappa\varepsilon_\nul(1+w)$ & $2H+\dfrac{{^3\!}R_\nul}{6H}$ \\
 & & & \\
  $\dfrac{-\,{^3\!}R_\nul}{{^3\!}R_\nul+3\kappa\varepsilon_\nul(1+w)}$ & $0$ &
     $\dfrac{6\varepsilon_\nul H(1+w)}{{^3\!}R_\nul+3\kappa\varepsilon_\nul(1+w)}$ &    $\dfrac{\tfrac{3}{2}\varepsilon_\nul(1+w)}{{^3\!}R_\nul+3\kappa\varepsilon_\nul(1+w)}$
     \end{tabular}
   \end{ruledtabular}
  \end{minipage}
\end{table*}

In this appendix, the perturbation theory~(\ref{subeq:final}) is
derived.  Before undertaking this task, it is necessary to carry out
two preliminary steps.

Firstly, as $\theta^{\phys}_{\een}=0$, (\ref{theta-gi}), it is
unnecessary to consider the gauge-dependent quantity~$\theta_{\een}$.
Eliminating the quantity~$\theta_{\een}$ from the differential
equations~(\ref{FRW6})–(\ref{FRW4a}) using the algebraic constraint
equation~(\ref{con-sp-1}) yields a system of four first-order ordinary
differential equations suitable for computer algebra software:
\begin{subequations}
\label{subeq:pertub-gi}
\begin{align}
   \dot{\varepsilon}_\een & + 3H(\varepsilon_\een +p_\een) \nonumber \\
     & +\varepsilon_\nul(1 + w)\Bigl[\vartheta_\een +\frac{1}{2H}\left(
 \kappa\varepsilon_\een+\tfrac{1}{2}\,{^3\!}R_\een\right)\Bigr] = 0,
       \label{FRW4gi} \\
   \dot{n}_\een & + 3H n_\een +
       n_\nul \Bigl[\vartheta_\een +\frac{1}{2H}\left(\kappa\varepsilon_\een +
      \tfrac{1}{2}\,{^3\!}R_\een\right)\Bigr] = 0, \label{FRW4agi} \\
  \dot{\vartheta}_\een & + H(2-3\beta^2)\vartheta_\een +
        \frac{1}{\varepsilon_\nul(1+w)}
      \dfrac{\tilde{\nabla}^2p_\een}{a^2} = 0,
\label{FRW5gi}\\
  {^3\!}\dot{R}_\een & +2H\,{^3\!}R_\een \nonumber \\
      &- 2\kappa\varepsilon_\nul(1 + w)\vartheta_\een
     +\frac{{^3\!}R_\nul}{3H} \left(\kappa\varepsilon_\een +
 \tfrac{1}{2}\,{^3\!}R_\een \right)  =0.  \label{FRW6gi} 
\end{align}
\end{subequations}
Secondly, the energy density and particle number density
perturbations~(\ref{e-n-gi}) are expressed in the four quantities
$\varepsilon_\een$, $n_\een$, $\vartheta_\een$, and ${^3\!}R_\een$.
Using the background equations~(\ref{subeq:einstein-flrw}) to
eliminate all time derivatives, and the linearized constraint
equation~(\ref{con-sp-1}) to eliminate~$\theta_\een$, the following
results are obtained:
\begin{subequations}
\label{subeq:pertub-gi-e-n}
\begin{align}
   & \varepsilon^\phys_\een  =
     \dfrac{ \varepsilon_\een{^3\!}R_\nul -
   3\varepsilon_\nul(1 + w) (2H\vartheta_\een +
  \tfrac{1}{2}\,{^3\!}R_\een) }
  { {^3\!}R_\nul+3\kappa\varepsilon_\nul(1 + w)},
        \label{Egi} \\
   & n^\phys_\een  = n_\een-\dfrac{3n_\nul(\kappa\varepsilon_\een+2H\vartheta_\een+
            \tfrac{1}{2}\,{^3\!}R_\een)}
         {{^3\!}R_\nul+3\kappa\varepsilon_\nul(1+w)}.  \label{nu2}
\end{align}
\end{subequations}
From an algebraic standpoint, it is more straightforward to begin by
deriving equations for $\varepsilon^\phys_\een$ and $n^\phys_\een$.
This will be accomplished in Sec.\,\ref{app:deriv-13}\@.
Subsequently, Eqs.\,(\ref{subeq:final}) for the relative
perturbations~(\ref{eq:contrast}) will be derived in
Sec.\,\ref{app:contrast}\@.  The expression~(\ref{eq:time-w}) will be
used to eliminate the time derivative of
$w:=p_{\nul}/\varepsilon_{\nul}$.

\subsection{Evolution Equation for Energy Density Perturbations}
\label{app:deriv-13}

The system~(\ref{subeq:pertub-gi}) and the expression~(\ref{Egi}) are
reformulated using~(\ref{eq:p1-a}) in a form suitable for
implementation in a computer algebra program:
\begin{subequations}
\label{subeq:nieuw}
\begin{align}
  \dot{\varepsilon}_\een+\alpha_{11}\varepsilon_\een+
    \alpha_{12}n_\een+
   \alpha_{13}\vartheta_\een+\alpha_{14}\,{^3\!}R_\een & = 0, \!
\label{nieuw1} \\
   \dot{n}_\een+\alpha_{21}\varepsilon_\een +
    \alpha_{22}n_\een+
   \alpha_{23}\vartheta_\een+\alpha_{24}\,{^3\!}R_\een & = 0, \!
\label{nieuw2} \\
  \dot{\vartheta}_\een +\alpha_{31}\varepsilon_\een +
     \alpha_{32}n_\een+
   \alpha_{33}\vartheta_\een+\alpha_{34}\,{^3\!}R_\een & = 0, \!
\label{nieuw3} \\
    {^3\!}\dot{R}_\een+\alpha_{41}\varepsilon_\een+
     \alpha_{42}n_\een+
   \alpha_{43}\vartheta_\een+\alpha_{44}\,{^3\!}R_\een & = 0, \!
\label{nieuw4} \\
  \varepsilon^\phys_\een+\alpha_{51}\varepsilon_\een+\alpha_{52}n_\een+
    \alpha_{53}\vartheta_\een+\alpha_{54}\,{^3\!}R_\een & = 0. \!
\label{nieuw5}
\end{align}
\end{subequations}
The coefficients $\alpha_{ij}$ are given in Table~\ref{eq:aij}.

The quantities in question~(\ref{subeq:pertub-gi-e-n}) do not include
the gauge function~$\psi(\bm{x})$. Consequently, the gauge
modes~(\ref{subeq:gauge-dep}) will naturally disappear as a result of
deriving the evolution equations for $\varepsilon^{\phys}_{\een}$ and
$n^{\phys}_{\een}$. This process will be carried out in three steps.

\paragraph*{Step 1.}
The first step is to remove the explicit occurrence of
${^3\!}R_{\een}$ from Eqs.\,(\ref{subeq:nieuw}).  By differentiating
Eq.\,(\ref{nieuw5}) with respect to time and eliminating the time
derivatives of $\varepsilon_{\een}$, $n_{\een}$, $\vartheta_{\een}$,
and ${^3\!}R_{\een}$ with the help of
Eqs.\,(\ref{nieuw1})--(\ref{nieuw4}), one obtains the following
equation:
\begin{equation}\label{eq:equiv}
   \dot{\varepsilon}^\phys_\een + p_1\varepsilon_\een+p_2 n_\een+
   p_3\vartheta_\een+p_4 \,{^3\!}R_\een=0,
\end{equation}
where the coefficients $p_1,\ldots,p_4$ are given by
\begin{equation}
  \label{eq:coef-pi}
  p_i=\dot{\alpha}_{5i}-\alpha_{51}\alpha_{1i}-
  \alpha_{52}\alpha_{2i}-\alpha_{53}\alpha_{3i}-\alpha_{54}\alpha_{4i}.
\end{equation}
As can be seen from Eq.\,(\ref{eq:equiv}), it follows that
\begin{equation}\label{eq:sol-3R1}
   {^3\!}R_\een=-\dfrac{1}{p_4}\dot{\varepsilon}^\phys_\een-
     \dfrac{p_1}{p_4}\varepsilon_\een-\dfrac{p_2}{p_4}n_\een-
     \dfrac{p_3}{p_4}\vartheta_\een.
\end{equation}
In this manner, ${^3\!}R_\een$ is expressed as a linear combination of
the quantities $\dot{\varepsilon}^\phys_\een$, $\varepsilon_\een$,
$n_\een$, and $\vartheta_\een$. Upon replacing ${^3\!}R_\een$ in
Eqs.\,(\ref{subeq:nieuw}) by the right-hand side of~(\ref{eq:sol-3R1}),
the following system of equations is obtained:
\begin{subequations}
\label{subeq:tweede}
\begin{align}
 \dot{\varepsilon}_\een+q_1\dot{\varepsilon}_\een^\phys+
   \gamma_{11}\varepsilon_\een+\gamma_{12}n_\een+
   \gamma_{13}\vartheta_\een & = 0, \label{tweede1} \\
 \dot{n}_\een+q_2\dot{\varepsilon}_\een^\phys+
    \gamma_{21}\varepsilon_\een+\gamma_{22}n_\een+
    \gamma_{23}\vartheta_\een & = 0, \label{tweede2} \\
 \dot{\vartheta}_\een+q_3\dot{\varepsilon}_\een^\phys+
   \gamma_{31}\varepsilon_\een+\gamma_{32}n_\een+
   \gamma_{33}\vartheta_\een & = 0, \label{tweede3} \\
     {^3\!}\dot{R}_\een+
   q_4\dot{\varepsilon}^\phys_\een+\gamma_{41}\varepsilon_\een+\gamma_{42}n_\een+
   \gamma_{43}\vartheta_\een & = 0, \label{tweede4} \\
 \varepsilon^\phys_\een+
   q_5\dot{\varepsilon}^\phys_\een+\gamma_{51}\varepsilon_\een+\gamma_{52}n_\een+
   \gamma_{53}\vartheta_\een & = 0, \label{tweede5}
\end{align}
\end{subequations}
where the coefficients $q_i$ and $\gamma_{ij}$ are given by
\begin{equation}\label{eq:betaij}
  q_i=-\dfrac{\alpha_{i4}}{p_4}, \quad
   \gamma_{ij}=\alpha_{ij}+q_i p_j.
\end{equation}
It has been achieved that ${^3\!}R_\een$ occurs explicitly only in
Eq.\,(\ref{tweede4}), whereas ${^3\!}R_\een$ occurs implicitly in the
remaining equations.  Therefore, Eq.\,(\ref{tweede4}) is no longer
required.  Accordingly, the remaining four ordinary differential
equations are as follows: (\ref{tweede1})--(\ref{tweede3})
and~(\ref{tweede5}) for the four unknown quantities
$\varepsilon_\een$, $n_\een$, $\vartheta_\een$,
and~$\varepsilon^\phys_\een$.

\paragraph*{Step 2.}
In a manner analogous to the approach undertaken in Step~1, the
explicit occurrence of $\vartheta_{\een}$ is removed from the system of
equations~(\ref{subeq:tweede}).  Differentiating Eq.\,(\ref{tweede5})
with respect to time and subsequently eliminating the time derivatives
of $\varepsilon_{\een}$, $n_\een$ and $\vartheta_\een$ with the help
of Eqs.\,(\ref{tweede1})--(\ref{tweede3}), results in the following
equation:
\begin{equation}
\label{eq:ddot-egi}
  q_5\ddot{\varepsilon}^\phys_\een+r\dot{\varepsilon}^\phys_\een+
     s_1\varepsilon_\een+s_2n_\een+s_3\vartheta_\een=0,
\end{equation}
where the coefficients $r$ and $s_i$ are given by
\begin{subequations}
\label{eq:coef-qi}
\begin{align}
  s_i & = \dot{\gamma}_{5i}-\gamma_{51}\gamma_{1i}-\gamma_{52}\gamma_{2i}-
       \gamma_{53}\gamma_{3i}, \\
  r & = 1+\dot{q}_5-\gamma_{51}q_1-\gamma_{52}q_2-\gamma_{53}q_3.
\end{align}
\end{subequations}
As can be seen from Eq.\,(\ref{eq:ddot-egi}), it follows that
\begin{equation}
\label{eq:sol-theta1}
  \vartheta\een=-\dfrac{q_5}{s_3}\ddot{\varepsilon}^\phys_\een-
     \dfrac{r}{s_3}\dot{\varepsilon}^\phys_\een-
     \dfrac{s_1}{s_3}\varepsilon_\een-\dfrac{s_2}{s_3}n_\een.
\end{equation}
In this manner, $\vartheta_\een$ is expressed as a linear combination
of $\ddot{\varepsilon}^\phys_\een$, $\dot{\varepsilon}^\phys_\een$,
$\varepsilon_\een$ and $n_\een$.  Upon replacing $\vartheta_\een$ in
Eqs.\,(\ref{subeq:tweede}) by the right-hand side
of~(\ref{eq:sol-theta1}), the follwing system of equations is
obtained:
\begin{subequations}
\label{subeq:derde}
\begin{align}
\dot{\varepsilon}_\een\,-&\;\gamma_{13}\dfrac{q_5}{s_3}\ddot{\varepsilon}^\phys_\een+
   \left(q_1-\gamma_{13}\dfrac{r}{s_3}\right)\dot{\varepsilon}^\phys_\een \nonumber\\
  +&\left(\gamma_{11}-\gamma_{13}\dfrac{s_1}{s_3}\right)\varepsilon_\een
  +\left(\gamma_{12}-\gamma_{13}\dfrac{s_2}{s_3}\right)n_\een  =0, \label{derde1}
\\
\dot{n}_\een\,-&\;\gamma_{23}\dfrac{q_5}{s_3}\ddot{\varepsilon}^\phys_\een+
   \left(q_2-\gamma_{23}\dfrac{r}{s_3}\right)\dot{\varepsilon}^\phys_\een\nonumber\\
    +&\left(\gamma_{21}-\gamma_{23}\dfrac{s_1}{s_3}\right)\varepsilon_\een
  +\left(\gamma_{22}-\gamma_{23}\dfrac{s_2}{s_3}\right)n_\een  =0, \label{derde2}
\\
\dot{\vartheta}_\een\,-&\;\gamma_{33}\dfrac{q_5}{s_3}\ddot{\varepsilon}^\phys_\een+
   \left(q_3-\gamma_{33}\dfrac{r}{s_3}\right)\dot{\varepsilon}^\phys_\een\nonumber\\
   +&\left(\gamma_{31}-\gamma_{33}\dfrac{s_1}{s_3}\right)\varepsilon_\een
  +\left(\gamma_{32}-\gamma_{33}\dfrac{s_2}{s_3}\right)n_\een  =0, \label{derde3}
\\
  {^3\!}\dot{R}_\een\,-&\;\gamma_{43}\dfrac{q_5}{s_3}\ddot{\varepsilon}^\phys_\een+
   \left(q_4-\gamma_{43}\dfrac{r}{s_3}\right)\dot{\varepsilon}^\phys_\een\nonumber\\
  +& \left(\gamma_{41}-\gamma_{43}\dfrac{s_1}{s_3}\right)\varepsilon_\een
  +\left(\gamma_{42}-\gamma_{43}\dfrac{s_2}{s_3}\right)n_\een  =0, \label{derde4}
\\
\varepsilon^\phys_\een\,-&\;\gamma_{53}\dfrac{q_5}{s_3}\ddot{\varepsilon}^\phys_\een+
   \left(q_5-\gamma_{53}\dfrac{r}{s_3}\right)\dot{\varepsilon}^\phys_\een\nonumber\\
    +&\left(\gamma_{51}-\gamma_{53}\dfrac{s_1}{s_3}\right)\varepsilon_\een
  +\left(\gamma_{52}-\gamma_{53}\dfrac{s_2}{s_3}\right)n_\een  =0. \label{derde5}
\end{align}
\end{subequations}
It has been achieved that the quantities $\vartheta_\een$ and
${^3\!}R_\een$ occur explicitly only in Eqs.\,(\ref{derde3})
and~(\ref{derde4}), whereas they occur implicitly in the remaining
equations.  Consequently, Eqs.\,(\ref{derde3}) and~(\ref{derde4}) are
no longer required.  The remaining
equations~(\ref{derde1}),~(\ref{derde2}) and~(\ref{derde5}) are three
ordinary differential equations for the three unknown quantities
$\varepsilon_\een$, $n_\een$, and $\varepsilon^\phys_\een$.

\paragraph*{Step 3.}
The subsequent steps would be to eliminate, in sequence,
$\varepsilon_\een$ and $n_\een$ from Eq.\,(\ref{derde5}) with the aid
of Eqs.\,(\ref{derde1}) and~(\ref{derde2}).  This would result in a
fourth-order differential equation for $\varepsilon^\phys_\een$.  It
must be noted, however, that this is not a possibility, given that the
gauge-dependent quantities $\varepsilon_\een$ and $n_\een$ do not
occur explicitly in Eq.\,(\ref{derde5}).  To demonstrate this,
Eq.\,(\ref{derde5}) will be rewritten in the subsequent form:
\begin{equation}
  \label{eq:eindelijk-nog-niet}
  \ddot{\varepsilon}^\phys_\een+a_1\dot{\varepsilon}^\phys_\een+
    a_2\varepsilon^\phys_\een=
    a_3\left(n_\een+\dfrac{\gamma_{51}s_3-\gamma_{53}s_1}
    {\gamma_{52}s_3-\gamma_{53}s_2} \varepsilon_\een\right),
\end{equation}
As a consequence of the gauge-invariance of the quantity
$\varepsilon^{\phys}_{\een}$, the left-hand side of
Eq.\,(\ref{eq:eindelijk-nog-niet}) is also gauge-invariant.  It thus
follows that the right-hand side of this equation is also
gauge-invariant, as will be demonstrated subsequently.  Indeed, the
result is as follows:
\begin{equation}\label{eq:check}
  \dfrac{\gamma_{51}s_3-\gamma_{53}s_1}{\gamma_{52}s_3-\gamma_{53}s_2}=
      -\dfrac{n_\nul}{\varepsilon_\nul(1+w)}.
\end{equation}
The proof of this equality is straightforward but requires significant
computational effort.  The algebraic task was performed using the
computer algebra system \textsc{Maxima}~\cite{maxima}.  The
conservation laws~(\ref{FRW2}) and~(\ref{FRW2a}) and the
expressions~(\ref{e-n-gi}) yield:
\begin{equation}
  \label{eq:equal-gi-non-gi}
  n_\een-\dfrac{n_\nul}{\varepsilon_\nul(1+w)}\varepsilon_\een=
  n^\phys_\een-\dfrac{n_\nul}{\varepsilon_\nul(1+w)}\varepsilon^\phys_\een.
\end{equation}
Substituting~(\ref{eq:check}) into~(\ref{eq:eindelijk-nog-niet}) and
using~(\ref{eq:equal-gi-non-gi}), we arrive at the evolution equation
for $\varepsilon^\phys_\een$:
\begin{equation}
  \label{eq:eindelijk}
  \ddot{\varepsilon}^\phys_\een+a_1\dot{\varepsilon}^\phys_\een+
  a_2\varepsilon^\phys_\een=a_{3}\left(n^\phys_\een-
    \dfrac{n_\nul}{\varepsilon_\nul(1+w)}\varepsilon^\phys_\een\right),
\end{equation}
where the coefficients $a_1$, $a_2$, and $a_3$ are given by
\begin{equation}
\label{subeq:vierde}
  a_1  = -\dfrac{s_3}{\gamma_{53}}+\dfrac{r}{q_5}, \quad
  a_2  = -\dfrac{s_3}{\gamma_{53}q_5}, \quad
  a_3  = \dfrac{\gamma_{52}s_3}{\gamma_{53}q_5}-\dfrac{s_2}{q_5}.
\end{equation}
The derivation of Eq.\,(\ref{eq:eindelijk}) shows that both the
perturbation ${^3\!}R_\een$ (\ref{driekrom}) of the spatial curvature
and the divergence $\vartheta_{\een}$ (\ref{fes5}) of the fluid
three-vector are included in this equation.

The final step in this process will be deriving an evolution equation
for~$n^\phys_\een$.  The linearized equations~(\ref{subeq:pertub-gi}),
from which the evolution equations are derived, are of fourth order.
From this system, a second-order equation~(\ref{eq:eindelijk}) for
$\varepsilon^\phys_\een$ has been derived.  It thus follows that the
remaining system (\ref{derde1})--(\ref{derde2}), and~(\ref{derde5})
from which an evolution equation for $n^\phys_\een$ can be derived is
at most of second order.  As the gauge-invariant quantities, namely
$\varepsilon^\phys_\een$ and $n^\phys_\een$, have been used, one
degree of freedom, specifically the gauge function $\psi(\bm{x})$
in~(\ref{subeq:gauge-dep}), has been eliminated.  Consequently, only a
first-order evolution equation for $n^\phys_\een$ can be derived.  In
place of deriving an equation for $n^\phys_\een$, an evolution
equation will be derived for the right-hand side of
expression~(\ref{eq:equal-gi-non-gi}).  By differentiating the
left-hand side of this expression with respect to time and using the
background equations~(\ref{FRW2}) and~(\ref{FRW2a}), the first-order
equations~(\ref{FRW4gi}) and~(\ref{FRW4agi}) and the definitions
$w:= p_\nul/\varepsilon_\nul$ and
$\beta^2:=\dot{p}_\nul/\dot{\varepsilon}_\nul$, the following result
is obtained:
\begin{align}
  \label{eq:vondst-gauge}
    \dfrac{1}{c} & \dfrac{\dif}{\dif t}\left(n_\een -
  \frac{n_\nul}{\varepsilon_\nul(1+w)}\varepsilon_\een\right)\nonumber\\
    &=-3H\left(1-\frac{n_\nul
  p_n}{\varepsilon_\nul(1+w)}\right) \left(n_\een -
  \frac{n_\nul}{\varepsilon_\nul(1+w)}\varepsilon_\een\right).
\end{align}
The algebraic task was performed using the
\textsc{Maxima}~\cite{maxima} computer algebra software.  Due to the
equality in Eq.\,(\ref{eq:equal-gi-non-gi}), it is possible to replace
$n_\een$ and $\varepsilon_\een$ in Eq.\,(\ref{eq:vondst-gauge}) by
$n^\phys_\een$ and~$\varepsilon^\phys_\een$.

\subsection{Perturbation Theory}
\label{app:contrast}

First Eq.\,(\ref{sec-ord}) will be derived. Substituting the expression
$\varepsilon^\phys_\een = \varepsilon_\nul\delta_\varepsilon$ into
Eq.\,(\ref{eq:eindelijk}) and dividing by $\varepsilon_\nul$ yields the
following result:
\begin{align}
 b_1=2\dfrac{\dot{\varepsilon}_\nul}{\varepsilon_\nul}+a_1, \quad
 b_2=\dfrac{\ddot{\varepsilon}_\nul}{\varepsilon_\nul}+
      a_1\dfrac{\dot{\varepsilon}_\nul}{\varepsilon_\nul}+a_2, \quad
 b_3=a_3\dfrac{n_\nul}{\varepsilon_\nul}.
\end{align}
The coefficients $a_{1}$, $a_{2}$, and $a_{3}$ are given
by~(\ref{subeq:vierde}).  With the aid of the computational software
\textsc{Maxima}~\cite{maxima}, the
coefficients~(\ref{subeq:coeff-contrast}) of the main text have been
calculated.

Finally, Eq.\,(\ref{fir-ord}) will be derived.  From the
definitions~(\ref{eq:contrast}), it can be deduced that
\begin{equation}\label{eq:sgi-contrast}
  n_\een^\phys-\frac{n_\nul}{\varepsilon_\nul(1+w)}\varepsilon_\een^\phys=
       n_\nul\left(\delta_n-\dfrac{\delta_\varepsilon}{1+w}\right).
\end{equation}
By using the equality~(\ref{eq:equal-gi-non-gi}) and substituting the
expression~(\ref{eq:sgi-contrast}) into Eq.\,(\ref{eq:vondst-gauge}),
the following is obtained:
\begin{align}\label{eq:diff-sgi}
  \dfrac{1}{c} & \dfrac{\dif}{\dif t}
   \left[n_\nul\left(\delta_n-\dfrac{\delta_\varepsilon}{1+w} \right)\right] \nonumber\\
&=-3H\left(1-\frac{n_\nul p_n}{\varepsilon_\nul(1+w)}\right)
\left[n_\nul\left(\delta_n-\dfrac{\delta_\varepsilon}{1+w} \right)\right].
\end{align}
Using Eq.\,(\ref{FRW2a}) yields Eq.\,(\ref{fir-ord}) of the main text.

\section{Verifying the Evolution Equations for Scalar Perturbations}
\label{app:consistency-check}

This article does not present a derivation of the linearized
equations~(\ref{subeq:basis}), from which the system of
equations~(\ref{subeq:pertub-flrw}) for scalar perturbations is
derived. Since the latter system is key to the cosmological
perturbation theory developed in this article, we will demonstrate its
validity.  This will be accomplished using three different methods.

The first step in this process is to apply a well-known property of
Einstein's equations and conservation laws.  This property stems from
the fact that the constraint equations contain only first-order time
derivatives of the metric tensor, while the dynamical equations
contain second-order time derivatives.  It implies that if the
constraint equations are satisfied at an initial time, then the
solutions to the dynamical equations and conservation laws will also
satisfy the constraint equations at all subsequent times.  A detailed
discussion can be found in Sec.\,7.5 of Weinberg's textbook~\cite{c8}
on the Cauchy problem.  The reverse is also true: if solutions to the
constraint equations and conservation laws can be found for all times,
then they necessarily satisfy the dynamical equations.  We will now
use this latter property to validate the system of
equations~(\ref{subeq:pertub-flrw}) for scalar perturbations.

The system~(\ref{subeq:pertub-flrw}) encompasses the conservation laws
and constraint equations.  It consists of four first-order ordinary
differential equations and one algebraic equation.  Therefore, it can
be solved, and its solutions automatically satisfy the system of
dynamical equations.  This fact will be used to prove the correctness
of Eqs.\,(\ref{subeq:pertub-flrw}).

For $i\neq j$, the dynamical equations~(\ref{basis-3}) are not coupled
to the scalar perturbations.  For the case of $i=j$, it is sufficient
to consider the contraction of~(\ref{basis-3}):
\begin{equation}
  \label{eq:contract-basis-3}
  \ddot{h}^{k}{}_{k}+6H\dot{h}^{k}{}_{k}+2\,{^3\!}R_\een=
     -3\kappa(\varepsilon_{\een}-p_{\een}).
\end{equation}
Using~(\ref{fes5}), yields
\begin{equation}
   (\dot{\theta}_\een-\dot{\vartheta}_\een) +
      6H(\theta_\een-\vartheta_\een)-{^3\!}R_\een=
       \tfrac{3}{2}\kappa(\varepsilon_\een-p_\een).
        \label{eq:168a}
\end{equation}
Eliminating the second term of equation~(\ref{eq:168a}) using the
energy density constraint equation~(\ref{con-sp-1}) results in
\begin{equation}
  \label{eq:168a-kort}
  \dot{\theta}_\een-\dot{\vartheta}_\een + \tfrac{1}{2}\,{^3\!}R_\een=
       -\tfrac{3}{2}\kappa(\varepsilon_\een+p_\een).
\end{equation}
As will now be shown, this equation is identical to the time
derivative of the energy density constraint equation~(\ref{con-sp-1}).
Differentiating equation~(\ref{con-sp-1}) with respect to time yields
\begin{equation}
\label{eq:con-sp-1-sum-time}
2\dot{H}(\theta_\een-\vartheta_\een)+2H(\dot{\theta}_\een-\dot{\vartheta}_\een)
 =\tfrac{1}{2}\,{^3\!}\dot{R}_\een+\kappa\dot{\varepsilon}_\een.
\end{equation}
Eliminating the time derivatives $\dot{H}$, ${^3\!}\dot{R}_\een$, and
$\dot{\varepsilon}_\een$ with the help of~(\ref{FRW3})--(\ref{FRW2}),
the momentum constraint equation~(\ref{FRW6}) and the energy density
conservation law~(\ref{FRW4}) yields the dynamical
equation~(\ref{eq:168a-kort}).  Consequently,
Eqs.\,(\ref{con-sp-1})--(\ref{FRW4}) are correct.

We will now continue with the second method of verifying
Eqs.\,(\ref{subeq:pertub-flrw}).  Equations~(\ref{FRW5})
and~(\ref{FRW4a}) have not yet been verified by the above procedure.
This will now be addressed by using the gauge modes.
Equations~(\ref{subeq:pertub-flrw}) are invariant under the gauge
transformations~(\ref{func}) with the gauge functions
$\xi^\mu(t,\bm{x})$ given by~(\ref{eq:synchronous}). Furthermore, the
quantities $\varepsilon_\een$, $n_\een$, $\theta_\een$,
$\vartheta_\een$, and ${^3\!}R_\een$ are gauge-dependent. It is
therefore evident that the gauge modes
\begin{subequations}
\label{subeq:gauge-dep}
\begin{align}
     \hat{S}_\een & =  \psi\dot{S}_\nul, \quad S=\varepsilon,n,T,p,\theta, \label{scalar-ijk} \\
  \hat{\vartheta}_\een &  =-\frac{\tilde{\nabla}^2\psi}{a^2},\label{theta-ijk}\\
    {^3\!}\hat{R}_\een  &  = 4H\left(\frac{\tilde{\nabla}^2\psi}{a^2} -
       \tfrac{1}{2}\,{^3\!}R_\nul\psi\right),
            \label{drie-ijk}
\end{align}
\end{subequations}
must be solutions to Eqs.\,(\ref{subeq:pertub-flrw}).  The gauge
modes~(\ref{scalar-ijk}) follow from~(\ref{eq:gauge-problem}), with
the understanding that $\xi^{0}=\psi(\bm{x})$.  The gauge
mode~(\ref{theta-ijk}) is the covariant divergence
of~(\ref{eq:gauge-mode-upar}).  As shown above, the linearized
Friedmann equation~(\ref{con-sp-1}) is correct.  Therefore, it can be
used to derive the gauge mode~(\ref{drie-ijk}).  Substituting the
gauge modes (\ref{scalar-ijk})--(\ref{theta-ijk}) into the linearized
Friedmann equation results in
\begin{align}
  \label{eq:pertub-friedmann-ijk}
  {^3\!}\hat{R}_\een&=4H(\hat{\theta}_{\een}-\hat{\vartheta}_{\een})-
                      2\kappa\hat{\varepsilon}_{\een} \nonumber\\
  &=4H\left(\psi\dot{\theta}_{\nul}+
        \frac{\tilde{\nabla}^2\psi}{a^2}\right)-
       2\kappa\psi\dot{\varepsilon}_{\nul}.
\end{align}
Using that $\theta_{\nul}=3H$, the Friedmann equation~(\ref{FRW3}),
Eq.\,(\ref{FRW3a}), and the energy density conservation
law~(\ref{FRW2}) yields the gauge mode~(\ref{drie-ijk}).
Expressions~(\ref{subeq:gauge-dep}) imply that the quantities
$S_{\een}$, $\vartheta_{\een}$, and ${^3\!}R_{\een}$ are invariant
under \emph{spatial} gauge transformations, i.e., transformations for
which the condition $\psi(\bm{x})=0$ holds. However, they are not
invariant under space-time gauge transformations.

It will now be demonstrated that the gauge modes given
by~(\ref{subeq:gauge-dep}) are solutions to
Eqs.\,(\ref{FRW6})--(\ref{FRW4a}). The first step to doing so is
deriving the time derivatives of the gauge modes:
\begin{subequations}
  \label{eqs:hulp-verg}
  \begin{align}
   \dot{\hat{\vartheta}}_{\een} & =-2H\hat{\vartheta}_{\een}, \\
   {^3\!}\dot{\hat{R}}_{\een} & =-4(\dot{H}-2H^{2})
         \bigl[\hat{\vartheta}_{\een}+
         \tfrac{1}{2}\,{^3\!}R_{\nul}\psi\bigr],\\
  \dot{\hat{\varepsilon}}_{\een}&= \psi\ddot{\varepsilon}_{\nul}  =
       -3\varepsilon_{\nul}(1+w)\bigl[\dot{H}-3H^{2}(1+\beta^{2})\bigr]\psi,\\
    \dot{\hat{n}}_{\een}&=\psi\ddot{n}_{\nul}=
            -3n_{\nul}\bigl[\dot{H}-3H^{2}\bigr]\psi,
  \end{align}  
\end{subequations}
where Eqs.\,(\ref{FRW3a})--(\ref{FRW2a}) and Eq.\,(\ref{eq:time-w})
have been used.  The validation of the
Eqs.\,~(\ref{FRW6})--(\ref{FRW4a}) can now be conducted.

First, substitute the gauge modes $\hat{\theta}_{\een}$,
$\hat{\vartheta}_{\een}$, and ${^3\!}\hat{R}_{\een}$ into
Eq.\,(\ref{FRW6}), to obtain an equation from which the gauge function
$\psi$ disappears.  Next, dividing this equation by
$\hat{\vartheta}_{\een}$ and multiplying the result by $\tfrac{3}{2}H$
yields the time derivative of Eq.\,(\ref{FRW3}).  Second, substitute
the gauge modes $\hat{\varepsilon}_{\een}$, $\hat{\theta}_{\een}$, and
$\hat{p}_{\een}$ into Eq.\,(\ref{FRW4}) eliminates the term with
$\dot{H}$.  Next, dividing the resulting equation by~$\psi$ and using
the definition $\beta^{2}:=\dot{p}_{\nul}/\dot{\varepsilon}_{\nul}$
followed by the use of Eq.\,(\ref{FRW2}) yields an identity.  Third,
substitute the gauge modes $\hat{\vartheta}_{\een}$, and
$\hat{p}_{\een}$ into Eq.\,(\ref{FRW5}), and using the definition
$\beta^{2}:=\dot{p}_{\nul}/\dot{\varepsilon}_{\nul}$ and subsequently
using Eq.\,(\ref{FRW2}) one arrives at an identity.  Finally, by
substituting the gauge modes $\hat{n}_{\een}$ and
$\hat{\theta}_{\een}$ into Eq.\,(\ref{FRW4a}) the term with $\dot{H}$
vanishes.  Using Eq.\,(\ref{FRW2a}), one arrives at an identity.  It
has been demonstrated that the gauge modes~(\ref{subeq:gauge-dep})
satisfy Eqs.\,(\ref{FRW6})--(\ref{FRW4a}). Consequently, it can be
concluded that these equations are correct.  Given the established
validity of Eq.\,(\ref{con-sp-1}), it can be deduced that the
system~(\ref{subeq:pertub-flrw}) is accurate as well.

As a final check, it is observed in Appendix\;\ref{sec:stand-theory}
that for a flat \textsc{flrw} universe, the evolution
equations~(\ref{subeq:algemene-vergelijkingen}) for the
gauge-dependent relative
perturbation~$\delta:=\varepsilon_{\een}/\varepsilon_{\nul}$ are
obtained from Eqs.\,(\ref{subeq:pertub-flrw}). In fact, the well-known
evolution equations for~$\delta$ are obtained in the cases of a
radiation-dominated universe~(\ref{subeq:standard}) and a universe
after matter and radiation
decouple~(\ref{subeq:algemene-vergelijkingen-mat-dom}).

Consequently, it can be concluded that the system of
equations~(\ref{subeq:pertub-flrw}) has been verified in three
distinct ways.

\section{Conventional Evolution Equation for Relative Density
  Perturbations}
\label{sec:stand-theory}

In this appendix the usual perturbation equations for relative
perturbations in a flat, ${^3\!}R_\nul=0$, \textsc{flrw} universe is
compared with the perturbation theory developed in this article. To
that end, the evolution equation for the relative perturbation
$\delta:=\varepsilon_\een/\varepsilon_\nul$ is derived using the
systems of background equations~(\ref{subeq:einstein-flrw-flat}) and
evolution equations~(\ref{subeq:pertub-gi}) for scalar
perturbations. In this appendix, we assume $c=1$ and $\Lambda=0$.

In the existing literature, the particle number density is usually not
taken into account.  Therefore, we take the equation of state for the
pressure $p=p(\varepsilon)$, so that ${p_{n}=0}$.  From~(\ref{eq:p1})
and~(\ref{eq:beta-matter}) we find that
$p_{\een}=p_{\varepsilon}\varepsilon_{\een}=\beta^{2}\varepsilon_{\een}$.
Using the definition $\delta:=\varepsilon_\een/\varepsilon_\nul$,
Eqs.\,(\ref{subeq:pertub-gi}) can be rewritten as follows:
\begin{subequations}
\label{subeq:pertub-gauge-dep}
\begin{align}
 & \dot{\delta} + 3H\delta\bigl(\beta^2+\tfrac{1}{2}(1-w)\bigr)+
     (1 + w)\left(\vartheta_\een + \dfrac{{^3\!}R_\een}{4H}\right)=0, \label{eq:dot-delta} \\
 & \dot{\vartheta}_\een + H(2-3\beta^2)\vartheta_\een +
        \frac{\beta^2}{1+w}
      \dfrac{\nabla^2\delta}{a^2} = 0, \label{FRW5gi-delta}\\
 & {^3\!}\dot{R}_\een + 2H\,{^3\!}R_\een - 
     2\kappa\varepsilon_\nul(1 + w)\vartheta_\een  =0, \label{FRW6gi-delta}
\end{align}
\end{subequations}
where Eq.\,(\ref{eq:energy-cons-law-flat}) has been used.  First,
differentiate equation~(\ref{eq:dot-delta}) with respect to time.
Then, eliminate the time derivatives of $w$, $H$, $\varepsilon_\nul$,
$\vartheta_\een$ and ${^3\!}R_\een$ using equations~(\ref{eq:time-w}),
(\ref{eq:energy-density-constraint-flat})
and~(\ref{eq:energy-cons-law-flat}) and the perturbation
equations~(\ref{FRW5gi-delta}) and~(\ref{FRW6gi-delta}).  Finally,
eliminate ${^3\!}R_\een$ using Eq.\,(\ref{eq:dot-delta}).  This
results in the system of equations:
\begin{subequations}
\label{subeq:algemene-vergelijkingen}
\begin{align}
  & \ddot{\delta} +2H\dot{\delta}\bigl(1+3\beta^2-3w\bigr) \nonumber\\
  & \qquad-\biggl[\beta^2\dfrac{\nabla^2}{a^2}+
   \tfrac{1}{2}\kappa\varepsilon_\nul\Bigl((1+w)(1+3w) \Bigr. \Bigr. \nonumber \\
  &  \qquad    +\,4w-6w^2+12\beta^2w-4\beta^2-
      6\beta^4 \Bigr)-6\beta\dot{\beta}H\biggr]\delta\nonumber\\  
  &  \quad =-3H\beta^2(1+w)\vartheta_\een, \label{eq:algemeen-p-e} \\
  & \dot{\vartheta}_\een+H(2-3\beta^2)\vartheta_\een+
    \dfrac{\beta^2}{1+w}\dfrac{\nabla^2\delta}{a^2}=0.
      \label{eq:continuity-equation}
\end{align}
\end{subequations}
In this procedure the computer algebra program
\textsc{Maxima}~\cite{maxima} has been used.  The gauge
modes~(\ref{subeq:gauge-dep}) are for the
system~(\ref{subeq:algemene-vergelijkingen}) given~by
\begin{subequations}
  \label{eq:gauge-modes-standard}
\begin{align}
  & \hat{\delta}(t,\bm{x}):=
     \dfrac{\psi(\bm{x})\dot{\varepsilon}_\nul(t)}{\varepsilon_\nul(t)}=
     -3H(t)\psi(\bm{x})\bigl(1+w(t)\bigr), \label{eq:delta-e--gauge}  \\
  & \hat{\vartheta}_\een(t,\bm{x})=
    -\dfrac{\nabla^2\psi(\bm{x})}{a^2(t)}.  \label{eq:theta-1-gauge}
\end{align}
\end{subequations}
From the Friedmann equation, $3H^2 =\kappa\varepsilon_\nul$, it
follows that $H\neq0$ for $\varepsilon_{\nul}\neq0$. This implies that
the gauge mode $\hat{\delta}(t,\bm{x})$ is nonzero, regardless of the
scale of the density perturbation.  Consequently, the general solution
of the system~(\ref{subeq:algemene-vergelijkingen}) has no physical
meaning. Therefore, the system~(\ref{subeq:algemene-vergelijkingen})
cannot be used to study the evolution of density perturbations.

\subsection{Radiation-Dominated Era}
\label{subsec:rad-dom-ijk}

In this era, the pressure is given by a linear barotropic relativistic
equation of state $p=\tfrac{1}{3}\varepsilon$, so that
$\beta^2=\tfrac{1}{3}$ and ${w=\tfrac{1}{3}}$.  The standard equation
is given by
\begin{equation}
  \label{eq:standard-rad-dom}
  \ddot{\delta} + 2H\dot{\delta}-
  \left(\dfrac{1}{3}\frac{\nabla^2}{a^2}+
   \tfrac{4}{3}\kappa\varepsilon_\nul\right)\delta =0.
\end{equation}
In contrast, Eqs.\,(\ref{subeq:algemene-vergelijkingen}) result in the
system
\begin{subequations}
\label{subeq:standard}
\begin{align}
 & \ddot{\delta} + 2H\dot{\delta}-
  \left(\dfrac{1}{3}\frac{\nabla^2}{a^2}+
   \tfrac{4}{3}\kappa\varepsilon_\nul\right)
   \delta =-\tfrac{4}{3}H\vartheta_\een, 
        \label{eq:delta-standard-genrel}  \\
 &  \dot{\vartheta}_\een+H\vartheta_\een+
           \dfrac{1}{4}\dfrac{\nabla^2\delta}{a^2}=0.
  \label{eq:continuity}
\end{align}
\end{subequations}
The gauge modes~(\ref{eq:gauge-modes-standard}) are solutions to the
system~(\ref{subeq:standard}) for $w=\tfrac{1}{3}$, as can be verified
by substitution.

For large-scale perturbations $\nabla^2\delta^\phys\rightarrow0$, the
solution to the system~(\ref{subeq:standard}) is
\begin{subequations}
  \label{eq:cont-eq-sol-phys-gauge}
  \begin{align}
   & \delta(t,\bm{x}) =(c_1t - 2\psi(\bm{x}) t^{-1}) + \tfrac{9}{8}t^{1/2}, \label{eq:cont-eq-sol-phys-gauge-A} \\
   & \vartheta_\een(t,\bm{x}) =-\dfrac{\nabla^2\psi(\bm{x})}{a^2}+\tfrac{9}{8}t^{-1/2},
                     \label{eq:cont-eq-sol-phys-gauge-B}
  \end{align}
\end{subequations}
where it is used that $H=\tfrac{1}{2}t^{-1}$ and
$\kappa\varepsilon_\nul=\tfrac{3}{4}t^{-2}$.  The expression between
parentheses in~(\ref{eq:cont-eq-sol-phys-gauge-A}) is the solution to
the homogeneous part of Eq.\,(\ref{eq:delta-standard-genrel}).  The
particular solution $\tfrac{9}{8}t^{1/2}$ is a consequence of
$\vartheta^\phys_\een=\tfrac{9}{8}t^{-1/2}$.  The right-hand side of
Eq.\,(\ref{eq:delta-standard-genrel}) is not equal to zero, which
means that Eq.\,(\ref{eq:standard-rad-dom}) is incomplete. As a
result, it does not provide all physical solutions.  Since
$\nabla^2\delta^\phys$ could have been large for small-scale
perturbations, it may have had a large influence on
$\vartheta^{\phys}_\een$ and this could have, in turn, a major impact
on the evolution of $\delta^\phys$.  Because the quantity
$\vartheta_{\een}$ is incorporated into
Eq.\,(\ref{eq:delta-rad}), oscillating relative density
perturbations~(\ref{nu13}) with an \emph{increasing} amplitude are
obtained instead of the \emph{constant} amplitude that follows
from~(\ref{eq:standard-rad-dom}).

\subsection{Era after Decoupling of Matter and Radiation}
\label{app:era-after-decoupling}

The standard equation is given by
\begin{equation}
  \label{eq:standaard-after-dec}
  \ddot{\delta}+2H\dot{\delta}-\left(\beta^2\dfrac{\nabla^2}{a^2}+
    \tfrac{1}{2}\kappa\varepsilon_\nul\right)\delta=0,
\end{equation}
where $\beta$ is given by~(\ref{eq:para-beta}).  In this era, the
equation of state for the pressure is according to thermodynamics
given by~(\ref{state-mat}). Therefore, (\ref{eq:para-w})
and~(\ref{eq:para-beta}) imply that $w\approx\tfrac{3}{5}\beta^2\ll1$,
so that with $T_\nul\propto a^{-2}$ one obtains
$\dot{\beta}/\beta=-H$.  Using
Eq.\,(\ref{eq:energy-density-constraint-flat}) one can derive the
following result:
$6\beta\dot{\beta}H=-2\kappa\varepsilon_\nul\beta^2$. Upon
substituting the latter expression into~(\ref{eq:algemeen-p-e}) and
neglecting $w$ and $\beta^2$ with respect to the constants of order
one, the system~(\ref{subeq:algemene-vergelijkingen}) results in
\begin{subequations}
\label{subeq:algemene-vergelijkingen-mat-dom}
\begin{align}
  & \ddot{\delta}+2H\dot{\delta}-\left(\beta^2\dfrac{\nabla^2}{a^2}+
    \tfrac{1}{2}\kappa\varepsilon_\nul\right)\delta=
    -3H\beta^2\vartheta_\een, \label{eq:alg-mat-dom-delta} \\
 &  \dot{\vartheta}_\een+2H\vartheta_\een+
           \beta^2\dfrac{\nabla^2\delta}{a^2}=0. \label{eq:cont-decoupling}
\end{align}
\end{subequations}
The gauge modes~(\ref{eq:gauge-modes-standard}) are solutions to the
system~(\ref{subeq:algemene-vergelijkingen-mat-dom}) for $w\ll1$ and
$\nabla^2\psi=0$, as can be verified by substitution.  Consequently,
for the system~(\ref{subeq:algemene-vergelijkingen-mat-dom}) $\psi$ is
an arbitrary infinitesimal constant $C$. This implies that
$\vartheta_\een=\vartheta^\phys_\een$ is a physical quantity, since
its gauge mode
$\hat{\vartheta}_\een$,~(\ref{eq:theta-1-gauge}), vanishes
identically.  Consequently, the relativistic gauge
transformation~(\ref{func}) with $\xi^\mu(t,\bm{x})$ given
by~(\ref{eq:synchronous}) reduces to the Newtonian gauge
transformation~(\ref{eq:gauge-trans-newt}).  This is to be expected,
since a cosmological fluid for which ${w\ll1}$ and $\beta^2\ll1$ can
be described by nonrelativistic equations of state~(\ref{state-mat}).

For large-scale perturbations $\nabla^2\delta^\phys\rightarrow0$, the
solution to the system~(\ref{subeq:algemene-vergelijkingen-mat-dom})
is
\begin{subequations}
  \label{subeq:sol-74ab}
\begin{align}
  & \delta(t) =(c_1t^{2/3}-2Ct^{-1})+\tfrac{7}{9}t^{-5/3},
      \label{eq:sol-74c1c2} \\
   & \vartheta_\een^\phys(t)  =-\tfrac{7}{9}t^{-4/3}, \label{eq:sol-theta-phys}
\end{align}
\end{subequations}
where $H=\tfrac{2}{3}t^{-1}$,
$\kappa\varepsilon_\nul=\tfrac{4}{3}t^{-2}$, and
$\beta^2\propto t^{-4/3}$ has been used.  The latter proportionality
follows from~(\ref{eq:para-w}) and~(\ref{eq:para-beta}),
$T_\nul\propto a^{-2}$, (\ref{eq:T-nul-propto-a-2}), and
$a\propto t^{2/3}$, (\ref{eq:exact-sol-mat}). The expression between
parentheses in~(\ref{eq:sol-74c1c2}) is the solution to the
homogeneous part of Eq.\,(\ref{eq:alg-mat-dom-delta}). The particular
solution $\tfrac{7}{9}t^{-5/3}$ is a consequence of
$\vartheta^\phys_\een=-\tfrac{7}{9}t^{-4/3}$.

The right-hand side of Eq.\,(\ref{eq:alg-mat-dom-delta}) is nonzero
for all scales.  Therefore, Eq.\,(\ref{eq:standaard-after-dec}) is
incomplete and does not yield all physical solutions.

\end{appendix}

\end{document}